\documentclass[twocolumn]{aastex62}
\usepackage{float}
\usepackage{amsmath}
\usepackage{afterpage}

\shorttitle{An Isolated Stellar-Mass Black Hole}
\shortauthors{Sahu et al.}

\begin{document}

\newcommand{\Gaia}{{\it Gaia}}
\newcommand{\Hipp}{{\it Hipparcos}}        
\newcommand{\Hubble}{{\it Hubble Space Telescope}}
\newcommand{\HST}{{\it HST}}
\newcommand{\IUE}{{\it IUE}}
\newcommand{\JWST}{{\it JWST}}
\newcommand{\kms}{{\>\rm km\>s^{-1}}}
\newcommand{\Mlens}{M_{\rm lens}}
\newcommand{\Spitzer}{{\it Spitzer}}
\newcommand{\Lya}{Ly$\alpha$}
\newcommand{\Teff}{T_{\rm eff}}
\newcommand{\OB}{MOA-11-191\slash OGLE-11-462}
\newcommand{\phil}{\hbox{$\varphi_{\mathrm{LS}}$}}
\newcommand{\msun}{$M_\odot$}
\newcommand{\amplification}{{magnification}}
\newcommand{\amplified}{magnified}

\newcommand{\DL}{D_{\rm L}}
\newcommand{\DS}{D_{\rm S}}
\newcommand{\masyr}{\rm mas\, yr^{-1}}
\newcommand{\piE}{\pi_{\rm E}}
\newcommand{\piLS}{\pi_{\rm LS}}
\newcommand{\tE}{t_{\rm E}}
\newcommand{\thetaE}{\theta_{\rm E}}
\renewcommand{\vec}[1]{\pmb{#1}}

\received{January 28, 2022}
\revised{May 19, 2022}
\accepted{May 24, 2022}
\published{July 6, 2022}
\submitjournal{ApJ}


\title{{An Isolated Stellar-Mass Black Hole Detected Through Astrometric Microlensing}
\footnote{This research is based in part on observations made with the NASA/ESA {\it Hubble Space Telescope\/}, obtained from the Space Telescope Science Institute, which is operated by the Association of Universities for Research in Astronomy, Inc., under NASA contract NAS 5-26555.}}

\correspondingauthor{Kailash C. Sahu}
\email{ksahu@stsci.edu}


\author[0000-0001-6008-1955]{Kailash C. Sahu}
\altaffiliation{PLANET Collaboration, MiNDSTEp Consortium}
\affil{Space Telescope Science Institute, 3700 San Martin Drive, Baltimore, MD 21218, USA}

\author[0000-0003-2861-3995]{Jay Anderson}
\affil{Space Telescope Science Institute, 3700 San Martin Drive, Baltimore, MD 21218, USA}

\author{Stefano Casertano}
\affil{Space Telescope Science Institute, 3700 San Martin Drive, Baltimore, MD 21218, USA}

\author[0000-0003-1377-7145]{Howard E. Bond}
\affil{Department of Astronomy \& Astrophysics, The Pennsylvania
State University, University Park, PA 16802, USA}
\affil{Space Telescope Science Institute, 3700 San Martin Drive, Baltimore, MD 21218, USA}

\author[0000-0001-5207-5619]{Andrzej Udalski}
\altaffiliation{OGLE Collaboration}
\affil{Astronomical Observatory,
University of Warsaw,
Al.\ Ujazdowskie 4,
00-478 Warszawa, Poland}

\author[0000-0002-3202-0343]{Martin Dominik}
\altaffiliation{MiNDSTEp Consortium, RoboNet Collaboration}
\affil{University of St Andrews, Centre for Exoplanet Science, SUPA School of Physics \& Astronomy, North Haugh, St Andrews, KY16 9SS, UK}

\author[0000-0002-0882-7702]{Annalisa Calamida}
\affil{Space Telescope Science Institute, 3700 San Martin Drive, Baltimore, MD 21218, USA}

\author[0000-0003-3858-637X]{Andrea Bellini}
\affil{Space Telescope Science Institute, 3700 San Martin Drive, Baltimore, MD 21218, USA}

\author[0000-0002-1793-9968]{Thomas M. Brown}
\affil{Space Telescope Science Institute, 3700 San Martin Drive, Baltimore, MD 21218, USA}

\author[0000-0002-6577-2787]{Marina Rejkuba}
\affil{European Southern Observatory,
Karl-Schwarzshild-Stra\ss e 2,
85748 Garching bei M\"unchen,
Germany
}

\author{Varun Bajaj}
\affil{Space Telescope Science Institute, 3700 San Martin Drive, Baltimore, MD 21218, USA}

\author[0000-0001-8803-6769]{No\'e Kains}
\altaffiliation{RoboNet Collaboration}
\affil{Department of Physics \& Astronomy, Barnard College, Columbia University, 3009 Broadway, New York, NY 10027, USA}

\author[0000-0001-7113-2738]{Henry C. Ferguson}
\affil{Space Telescope Science Institute, 3700 San Martin Drive, Baltimore, MD 21218, USA}

\author[0000-0003-2624-0056]{Chris L. Fryer}
\affil{Center for Theoretical Astrophysics, Los Alamos National Laboratory, Los Alamos, NM 87545, USA}

\author[0000-0001-9716-7752]{Philip Yock}
\affil{Department of Physics, University of Auckland, Auckland, New Zealand}

\nocollaboration

\author{Przemek Mr{\'o}z}
\affil{Astronomical Observatory,
University of Warsaw,
Al.\ Ujazdowskie 4,
00-478 Warszawa, Poland}

\author{Szymon Koz{\l}owski}
\affil{Astronomical Observatory,
University of Warsaw,
Al.\ Ujazdowskie 4,
00-478 Warszawa, Poland}

\author{Pawe{\l} Pietrukowicz}
\affil{Astronomical Observatory,
University of Warsaw,
Al.\ Ujazdowskie 4,
00-478 Warszawa, Poland}

\author{Radek Poleski}
\affil{Astronomical Observatory,
University of Warsaw,
Al.\ Ujazdowskie 4,
00-478 Warszawa, Poland}

\author{Jan Skowron}
\affil{Astronomical Observatory,
University of Warsaw,
Al.\ Ujazdowskie 4,
00-478 Warszawa, Poland}

\author{Igor Soszy{\'n}ski}
\affil{Astronomical Observatory,
University of Warsaw,
Al.\ Ujazdowskie 4,
00-478 Warszawa, Poland}

\author{Micha\l \,K. Szyma{\'n}ski}
\affil{Astronomical Observatory,
University of Warsaw,
Al.\ Ujazdowskie 4,
00-478 Warszawa, Poland}

\author{Krzysztof Ulaczyk}
\affil{Astronomical Observatory,
University of Warsaw,
Al.\ Ujazdowskie 4,
00-478 Warszawa, Poland}
\affil{Department of Physics, University of Warwick, Gibbet Hill Road,
Coventry, CV4~7AL, UK}

\author{{\L}ukasz Wyrzykowski}
\affil{Astronomical Observatory,
University of Warsaw,
Al.\ Ujazdowskie 4,
00-478 Warszawa, Poland}

\collaboration{(OGLE Collaboration)}

\author{Richard Barry}
\affiliation{Code 667, NASA Goddard Space Flight Center, Greenbelt, MD 20771, USA}
\author{David P.~Bennett}
\affiliation{Code 667, NASA Goddard Space Flight Center, Greenbelt, MD 20771, USA}
\affiliation{Department of Astronomy, University of Maryland, College Park, MD 20742, USA}
\author{Ian A. Bond}
\affiliation{Institute of Natural and Mathematical Sciences, Massey University, Auckland 0745, New Zealand}
\author{Yuki Hirao}
\affiliation{Department of Earth and Space Science, Graduate School of Science, Osaka University, Toyonaka, Osaka 560-0043, Japan}
\author[0000-0003-2267-1246]{Stela {Ishitani~Silva}}
\affiliation{Code 667, NASA Goddard Space Flight Center, Greenbelt, MD 20771, USA}
\affiliation{Department of Physics, The Catholic University of America, Washington, DC 20064, USA}
\author{Iona Kondo}
\affiliation{Department of Earth and Space Science, Graduate School of Science, Osaka University, Toyonaka, Osaka 560-0043, Japan}
\author{Naoki Koshimoto}
\affiliation{Code 667, NASA Goddard Space Flight Center, Greenbelt, MD 20771, USA}
\author{Cl\'ement Ranc}
\affil{Zentrum f{\"u}r Astronomie der Universit{\"a}t Heidelberg, Astronomisches Rechen-Institut, M{\"o}nchhofstr.\ 12-14, 69120 Heidelberg, Germany}
\author{Nicholas J. Rattenbury}
\affiliation{Department of Physics, University of Auckland, Private Bag 92019, Auckland, New Zealand}
\author{Takahiro Sumi}
\affiliation{Department of Earth and Space Science, Graduate School of Science, Osaka University, Toyonaka, Osaka 560-0043, Japan}
\author{Daisuke Suzuki}
\affiliation{Department of Earth and Space Science, Graduate School of Science, Osaka University, Toyonaka, Osaka 560-0043, Japan}
\author{Paul J. Tristram}
\affiliation{University of Canterbury Mt.\ John Observatory, P.O. Box 56, Lake Tekapo 8770, New Zealand}
\author{Aikaterini Vandorou}
\affiliation{Code 667, NASA Goddard Space Flight Center, Greenbelt, MD 20771, USA}
\affiliation{Department of Astronomy, University of Maryland, College Park, MD 20742, USA}

\collaboration{(MOA collaboration)}

\author{Jean-Philippe~Beaulieu}
\affil{School of Natural Sciences, University of Tasmania, Private Bag 37 Hobart, Tasmania 7001 Australia}
\affil{Sorbonne Universite, UPMC Univ Paris 6 et CNRS, UMR 7095, Institut d'Astrophysique de Paris, 98 bis bd Arago, 75014 Paris, France}

\author{Jean-Baptiste~Marquette}
\affil{Laboratoire d'astrophysique de Bordeaux, Univ.\ Bordeaux, CNRS, B18N, all\'ee Geoffroy SaintHilaire, 33615 Pessac, France}

\author{Andrew Cole}
\affil{School of Natural Sciences, University of Tasmania, Private Bag 37 Hobart, Tasmania 7001 Australia}

\author{Pascal Fouqu{\'e}}
\affil{Université de Toulouse, UPS-OMP, IRAP, Toulouse, France}

\author{Kym Hill}
\affil{School of Natural Sciences, University of Tasmania, Private Bag 37 Hobart, Tasmania 7001 Australia}

\author{Stefan Dieters}
\affil{School of Natural Sciences, University of Tasmania, Private Bag 37 Hobart, Tasmania 7001 Australia}

\author{Christian Coutures}
\affil{Sorbonne Universite, UPMC Univ Paris 6 et CNRS, UMR 7095, Institut d'Astrophysique de Paris, 98 bis bd Arago, 75014 Paris, France}

\author{Dijana Dominis-Prester}
\affil{Okayama Astrophysical Observatory, National Astronomical Observatory of Japan, Asakuchi, Okayama 719-0232, Japan}

\author{Clara Bennett}
\affil{Department of Physics, Massachussets Institute of Technology, Cambridge, MA 02139, USA}

\author{Etienne Bachelet}
\affil{Las Cumbres Observatory Global Telescope Network, 6740 Cortona Drive, Suite 102, Goleta, CA 93117, USA}

\author{John Menzies}
\affil{South African Astronomical Observatory, PO Box 9, Observatory 7935, South Africa}

\author{Michael Albrow}
\affil{University of Canterbury, Department of Physics \& Astronomy, Private Bag 4800, Christchurch 8020, New Zealand}

\author{Karen Pollard}
\affil{University of Canterbury, Department of Physics \& Astronomy, Private Bag 4800, Christchurch 8020, New Zealand}

\collaboration{(PLANET Collaboration)}

\author{Andrew Gould}
\affil{Max-Planck-Institute for Astronomy, K\"onigstuhl 17, 69117 Heidelberg, Germany}
\affil{Department of Astronomy, Ohio State University, 140 W. 18th Ave., Columbus, OH 43210, USA}

\author{Jennifer C. Yee}
\affil{
Center for Astrophysics $|$ Harvard \& Smithsonian,
60 Garden St., Cambridge, MA 02138, USA}

\author{William Allen}
\affiliation{Vintage Lane Observatory, Blenheim, New Zealand}

\author{Leonardo A. Almeida}
\affiliation{
Escola de Ci\^encias e Tecnologia, Universidade Federal do Rio Grande do Norte, Natal - RN, 59072-970, Brazil}
\affiliation{Programa de P\'os-Gradua\c{c}\~ao em F\'isica, Universidade do Estado do Rio Grande do Norte, Mossor\'o - RN, 59610-210, Brazil}

\author{Grant Christie}
\affiliation{Auckland Observatory, Auckland, New Zealand}

\author{John Drummond}
\affiliation{Possum Observatory, Patutahi, New Zealand}
\affiliation{Centre for Astrophysics, University of Southern Queensland,
Toowoomba, Queensland 4350, Australia}

\author{Avishay Gal-Yam}
\affiliation{Benoziyo Center for Astrophysics, Weizmann Institute of Science, 76100 Rehovot, Israel}

\author{Evgeny Gorbikov}
\affiliation{School of Physics and Astronomy, Raymond and Beverley Sackler Faculty of Exact Sciences, Tel-Aviv University, Tel Aviv 69978, Israel}

\author{Francisco Jablonski}
\affiliation{
Instituto Nacional de Pesquisas Espaciais,
Astrophysics Division,
Sao Jose dos Campos, Brazil}

\author{Chung-Uk Lee}
\affiliation{Korea Astronomy and Space Science Institute, Daejon 34055, Republic of Korea}

\author{Dan Maoz}
\affiliation{School of Physics and Astronomy, Tel-Aviv University, Tel-Aviv 6997801, Israel}

\author{Ilan Manulis}
\affiliation{Department of Particle Physics and Astrophysics, Weizmann Institute of Science, 76100 Rehovot, Israel}

\author{Jennie McCormick}
\affiliation{Farm Cove Observatory, Centre for Backyard Astrophysics, Pakuranga, Auckland, New Zealand}

\author{Tim Natusch}
\affiliation{Auckland Observatory, Auckland, New Zealand}
\affiliation{Institute for Radio Astronomy and Space Research (IRASR), AUT University, Auckland, New Zealand}

\author{Richard W. Pogge}
\affiliation{Department of Astronomy, Ohio State University, 140 W. 18th Ave., Columbus, OH  43210, USA}

\author{Yossi Shvartzvald}
\affiliation{Department of Particle Physics and Astrophysics, Weizmann Institute of Science, Rehovot 76100, Israel}

\collaboration{($\mu$FUN Collaboration)}
\author{Uffe G. J{\o}rgensen}
\affil{Centre for ExoLife Sciences, Niels Bohr Institute, University of Copenhagen, {\O}ster Voldgade 5, 1350 Copenhagen, Denmark}

\author{Khalid A. Alsubai}
\affil{Qatar Environment and Energy Research Institute (QEERI), HBKU, Qatar Foundation, Doha, Qatar}

\author{Michael I. Andersen}
\affil{Niels Bohr Institute, University of Copenhagen, Blegdamsvej 17, 2100 Copenhagen, Denmark}

\author{Valerio Bozza}
\affil{Dipartimento di Fisica ``E.R. Caianiello,'' Universit{\`a} di Salerno, Via Giovanni Paolo II 132, 84084 Fisciano, Italy}
\affil{Istituto Nazionale di Fisica Nucleare, Sezione di Napoli, Napoli, Italy}

\author{Sebastiano Calchi Novati}
\affil{IPAC, Mail Code 100-22, Caltech, 1200 E. California Blvd., Pasadena, CA 91125, USA}

\author{Martin Burgdorf}
\affil{Lianenweg 7a, 22529 Hamburg, Germany}

\author{Tobias C. Hinse}
\affil{Institute of Astronomy, Faculty of Physics, Astronomy, and Informatics, Nicolaus Copernicus University in Toru\'n, ul.\ Grudziadzka 5, 87-100 Toru\'n, Poland}
\affil{Chungnam National University, Department of Astronomy, Space Science and Geology, Daejeon, South Korea}

\author{Markus Hundertmark}
\altaffiliation{RoboNet Collaboration}
\affil{Zentrum f{\"u}r Astronomie der Universit{\"a}t Heidelberg, Astronomisches Rechen-Institut, M{\"o}nchhofstr.\ 12-14, 69120 Heidelberg, Germany}

\author{Tim-Oliver Husser}
\affil{Institut fur Astrophysik, Georg-August-Universität G\"{o}ttingen, Friedrich-Hund-Platz 1, 37077 G\"{o}ttingen, Germany}

\author{Eamonn Kerins}
\affil{Jodrell Bank Centre for Astrophysics, Alan Turing Building, University of Manchester, Manchester, M13 9PL, UK}

\author{Penelope Longa-Pe{\~n}a}
\affil{Centro de Astronom{\'{\i}}a, Universidad de Antofagasta, Avenida Angamos 601, Antofagasta 1270300, Chile}

\author{Luigi Mancini}
\affil{Department of Physics, University of Rome ``Tor Vergata,'' Via della Ricerca Scientifica 1, 00133 Roma, Italy}
\affil{Max-Planck-Institute for Astronomy, K\"onigstuhl 17, 69117 Heidelberg, Germany}
\affil{INAF -- Astrophysical Observatory of Turin, Via Osservatorio 20, 10025 Pino Torinese, Italy}

\author{Matthew Penny}
\affil{Louisiana State University, 261-B Nicholson Hall, Tower Dr., Baton Rouge, LA 70803-4001, USA}

\author{Sohrab Rahvar}
\affil{Department of Physics, Sharif University of Technology, PO Box 11155-9161, Tehran, Iran}

\author{Davide Ricci}
\affil{INAF -- Padova Astronomical Observatory, Vicolo dell'Osservatorio 5, 35122 Padova, Italy}

\author{Sedighe Sajadian}
\affil{Department of Physics, Isfahan University of Technology, Isfahan 84156-83111, Iran }

\author{Jesper Skottfelt}
\affil{Centre for Electronic Imaging, Department of Physical Sciences, The Open University, Milton Keynes, MK7 6AA, UK}

\author{Colin Snodgrass}
\altaffiliation{RoboNet Collaboration}
\affil{Institute for Astronomy, University of Edinburgh, Royal Observatory, Edinburgh, EH9 3HJ, UK}

\author{John Southworth}
\affil{Astrophysics Group, Keele University, Staffordshire, ST5 5BG, UK}

\author{Jeremy Tregloan-Reed}
\affil{Instituto de Investigaci\'on en Astronom\'ia y Ciencias Planetarias, Universidad de Atacama, Copayapu 485, Copiap\'o, Atacama, Chile}

\author{Joachim Wambsganss}
\affil{Zentrum f{\"u}r Astronomie der Universit{\"a}t Heidelberg, Astronomisches Rechen-Institut, M{\"o}nchhofstr.\ 12-14, 69120 Heidelberg, Germany}

\author{Olivier Wertz}
\affil{Space Sciences, Technologies, and Astrophysics Research (STAR) Institute, University of Li\`{e}ge, Li\`{e}ge, Belgium}

\collaboration{(MiNDSTEp Consortium)}

\author{Yiannis Tsapras}
\affil{Zentrum f{\"u}r Astronomie der Universit{\"a}t Heidelberg, Astronomisches Rechen-Institut, M{\"o}nchhofstr.\ 12-14, 69120 Heidelberg, Germany}

\author{Rachel A. Street}
\affil{Las Cumbres Observatory Global Telescope Network, 6740 Cortona Drive, Suite 102, Goleta, CA 93117, USA}

\author{D. M. Bramich}
\affil{Center for Astro, Particle, and Planetary Physics, New York University Abu Dhabi, P.O. Box 129188, Saadiyat Island, Abu Dhabi, UAE}
\affil{Division of Engineering, New York University Abu Dhabi, P.O. Box 129188, Saadiyat Island, Abu Dhabi, UAE}

\author{Keith Horne}
\altaffiliation{PLANET Collaboration}
\affil{University of St Andrews, Centre for Exoplanet Science, SUPA School of Physics \& Astronomy, North Haugh, St Andrews, KY16 9SS, UK}

\author{Iain A. Steele}
\affil{Astrophysics Research Institute, Liverpool John Moores University, Liverpool, CH41 1LD, UK}

\collaboration{(RoboNet Collaboration)}




\begin{abstract}

We report the first unambiguous detection and mass measurement of an isolated
stellar-mass black hole (BH). We used the {\it Hubble Space Telescope\/} (\HST\/)
to carry out precise astrometry of the source star of the long-duration
($t_{\rm E} \simeq 270$~days), high-magnification microlensing event MOA-2011-BLG-191\slash  OGLE-2011-BLG-0462 (hereafter designated as MOA-11-191\slash OGLE-11-462), in the direction of the Galactic bulge. \HST\/ imaging, conducted at eight epochs over an interval of six years, reveals a clear relativistic astrometric deflection of the background star's apparent position.  Ground-based photometry of MOA-11-191\slash OGLE-11-462 shows a parallactic signature of the effect of the Earth's motion on the microlensing light curve.
Combining the \HST\/ astrometry with the ground-based light curve and the derived parallax, we obtain a lens mass of $7.1\pm1.3\,M_\odot$ and a distance of $1.58\pm0.18$~kpc. 
We show that the lens emits no detectable light, which, along with having a mass higher than is possible for a white dwarf or neutron star, confirms its BH nature. Our analysis also provides an absolute proper motion for the BH\null. The proper motion is offset from the mean motion of Galactic-disk stars at similar distances by an amount corresponding to a transverse space velocity of $\sim$$45\,\kms$, suggesting that the BH received a ``natal kick'' from its supernova explosion. Previous mass determinations for stellar-mass BHs have come from radial-velocity measurements of Galactic X-ray binaries, and from gravitational radiation emitted by merging BHs in binary systems in external galaxies. Our mass measurement is the first for an isolated stellar-mass BH using any technique.

\null\vskip 0.2in
\end{abstract}

\section{Measuring the Masses of Black Holes \label{section:measuringthemasses} }

\subsection{Black Holes in Binary Systems \label{subsec:bhinbinaries} }

Stars with initial masses greater than $\sim$$20\,M_\odot$ are expected to end their lives as black holes (BHs) \citep[e.g.,][]{Fryer2001,Woosley2002,Heger2003,Spera2015,Sukhbold2016}. Objects of these masses constitute roughly 0.1\% of all stars, leading to the expectation that the Galaxy should now contain of the order of $\approx$$10^8$ BHs \citep{Shapiro1983,vandenHeuvel1992,Brown1994,Samland1998}.

However, the actual detection of stellar-mass BHs is observationally challenging, and determining their masses even more so. BHs have been identified in the Galaxy and Local Group through X-ray emission due to accretion in short-period binary systems, most of them soft X-ray transients. In such cases, dynamical masses of the BHs can be measured or estimated through radial-velocity measurements and light-curve modeling for the optical companion stars (the techniques are reviewed by \citealt{Remillard2006} and \citealt{Casares2014}). Masses of nearly two dozen BHs in X-ray binary systems have been determined using these methods, with varying degrees of precision. These ``electromagnetically measured'' BH masses show a distribution peaking near 7--$8\,M_\odot$, with few if any below $\sim$$5\,M_\odot$ \citep[e.g.,][]{Ozel2010,Farr2011,Kreidberg2012,Corral-Santana2016}. This suggests that a ``mass gap'' exists between the lowest-mass BHs, and the highest measured masses of neutron stars (NSs) in binary radio pulsars of $\sim$2.1--$2.3\,M_\odot$ \citep{Linares2018,Cromartie2020}. 
Recently, however, a few non-accreting or weakly accreting BHs have been discovered in longer-period spectroscopic binaries in the field \citep[e.g.,][]{Thompson2019,Jayasinghe2021} and in globular clusters \citep{Giesers2019}, lying in the NS-BH gap with dynamical masses of $\sim$3--$4.5\,M_\odot$. 
Precision astrometry of nearby stars by \Gaia\/ shows the promise of detecting additional wide binary systems containing quiescent BHs \citep[e.g.,][and references therein]{Chawla2021,Janssens2021} and measuring their masses.
At the high-mass end, the BH mass distribution falls off above $\sim$$10\,M_\odot$, and very few electromagnetic BH masses are known above $\sim$$15\,M_\odot$, the only exception in the Milky Way being an updated mass determination of $21.1\pm 2.2\,M_\odot$ for the BH in Cygnus~X-1 \citep{Miller-Jones2021}. Among extragalactic X-ray binaries, BH masses as high as $15.65\pm1.45$ and $17\pm4\,M_\odot$ have been reported for M33 X-7 \citep{Orosz2007} and NGC\,300 X-1 \citep{Binder2021}, respectively. An even higher mass of at least $23.1\,M_\odot$ was reported for the compact object in IC\,10 X-1 \citep{Silverman2008}, but this has been questioned \citep{Laycock2015}.

The first detections of gravitational waves (GWs) by the Laser Interferometry Gravitational Wave Observatory (LIGO) and Virgo Collaboration (\citep{Abbott2016})  revealed a population of massive merging binary BHs, BH-NS pairs, and binary NSs at extragalactic distances. In the source catalogs from the third observing run of the Advanced LIGO and Advanced Virgo collaboration \citep{Abbott2021a,Abbott2021b}, the inferred masses of the BHs among the pre-merger systems range from $\approx$6 to $95\,M_\odot$, with two low-mass outliers among the secondary components at $\sim$2.6 and $2.8\,M_\odot$, which could be either BHs or NSs\null.

\subsection{Isolated Black Holes \label{subsec:isolatedbhs} }

The electromagnetic and GW mass measurements described above are all for BHs in binary systems, including those undergoing mass accretion or mergers. However, there are reasons to believe that a substantial fraction of stellar-mass BHs are single, rather than belonging to binaries. First, about 30\% of massive stars are born single \citep[see][]{Sana2012,deMink2014}. Second, in a close binary system, the pair may enter into a common envelope and merge before the supernova (SN) explosion \citep[e.g.,][]{Fryer1999, Zhang2001, Tutukov2011, Dominik2012}.
Lastly, in a wide binary, the ``natal kick'' imparted to the companion by the SN event may be large enough to detach the two components, producing an isolated BH  (e.g., \citealt{Tauris2006,Belczynski2016}).
Being less altered by interactions with companions, single BHs potentially provide a more direct probe of BH formation than those in binaries.

Isolated BHs are extremely difficult to detect directly. They emit no light of their own, and the accretion rate from the interstellar medium is generally likely to be too low to produce detectable X-ray or radio emission (see, however, \citealt{Agol2002,Fender2013,Tsuna2019,Scarcella2021} for the case of isolated BHs in dense environments). In fact, until now, no isolated stellar-mass BH has ever been unambiguously found within our Galaxy or elsewhere.

Microlensing is the only available method for measuring the masses of isolated BHs.
{\it Astrometric microlensing\/}---the relativistic deflection of the apparent position of a background star when a compact object passes in front of it---provides  a direct method for measuring the masses of BH lenses.  High spatial-resolution interferometric observations of microlensing events \citep{Dong2019, zang2020}, and observations of rare events where the lens passes over the surface of the source  \citep{Yoo2004}, can also yield the masses of BH lenses. In this paper, we describe how the technique of astrometric microlensing is used to determine masses. We discuss our ongoing program of astrometric measurements of microlensing events with the {\it Hubble Space Telescope\/} (\HST\/). Then we report the first detection of an isolated BH and our measurement of its mass. 

\section{Measuring the Masses of Isolated Black Holes with Astrometric Microlensing}

\subsection{{Microlensing Events and Black-Hole Candidates}}

A microlensing event occurs when a star or compact object (the lens) passes almost exactly in front of a background star (the source). As predicted by general relativity \citep{Einstein1936}, the lens magnifies the image of the source, producing an apparent amplification of its brightness. The lens also slightly shifts the apparent position of the source \citep{Miyamoto1995,Hog1995,Walker1995}---an analog of the deflection of stellar images during the 1919 solar eclipse \citep{Dyson1920}, which provided support for the general theory of relativity.

Microlensing survey programs, including OGLE \citep{Udalski2015}, MOA \citep{Bond2001}, 
and KMTNet \citep{Kim2016}, carry out photometric monitoring of rich stellar fields in the Galactic bulge. These surveys typically detect $>$2000 events toward the Galactic bulge annually. To date, more than 30,000 microlensing events have been discovered and monitored by these survey programs.

The characteristic scale of gravitational microlensing is provided by the angular Einstein radius $\theta_\mathrm{E}$, given by 
\begin{equation}
\theta_\mathrm{E} \equiv  \sqrt{\frac{4 G M_{\rm lens}}{c^2} \frac{\pi_\mathrm{LS}}{1~\mbox{AU}}} \, ,
\label{eq:theta}
\end{equation}
where $\Mlens$ is the mass of the lens, and 
\begin{equation}
\piLS \equiv \pi_\mathrm{L}-\pi_\mathrm{S} = (1~\mbox{AU})  \left(\frac{1}{D_\mathrm{L}} - \frac{1}{D_\mathrm{S}}\right) 
\label{eq:distances}
\end{equation}
is the relative lens-source parallax, with
$D_\mathrm{L}$ and $D_\mathrm{S}$ being the distances from the observer to the lens and to the source, respectively.

The Einstein radius $\theta_\mathrm{E}$, however, cannot be obtained directly from the magnification light curve,
whose only characteristic that carries a physical dimension is the timescale. Given the relative proper motion $\mu_\mathrm{LS}$ between lens and source, we can straightforwardly define 
$t_\mathrm{E}^\star$, the time for the source to  
traverse an angular distance of $\theta_\mathrm{E}$ in the barycentric
reference frame, as 
$t_\mathrm{E}^\star = \theta_\mathrm{E}/\mu_\mathrm{LS}$
(more details in \S\ref{sec:pardef}).
The distribution of the timescale $t_\mathrm{E}^\star$ of the observed events peaks around 25 days, with $t_\mathrm{E}^\star$ ranging from a fraction of a day to several hundred days \citep[e.g.,][]{Wyrzykowski2015}.

If BHs constitute a small
but non-negligible
fraction of the total stellar mass of the Galaxy, as described above, then a few of the observed microlensing events are expected to be due to BHs. 
Equation \ref{eq:theta} shows that the angular Einstein radius $\theta_\mathrm{E}$ is proportional to the square root of the lens mass. 
So, all else being equal, events due to massive compact objects would preferentially tend to be characterized by longer event durations ($\tE \gtrsim $ 150 days) combined with an apparent lack of light contribution from the lens.
However, 
a degeneracy between lens mass and proper motion remains.
Thus a long-duration event with no light contribution from the lens could arise from a high-mass, non-luminous  BH lens with a large Einstein radius---but it could alternatively be due simply to an unusually slow-moving, faint, low-mass ordinary star. If the lens were a luminous massive star, this would generally be recognizable through the contribution of its light, particularly when observations are available in two different bandpasses.

Indeed, several long-duration OGLE and MOA microlensing events 
have been suggested as being due to BHs  \citep[e.g.,][]{Bennett2002,Mao2002,Minniti2015,Wyrzykowski2020}.  However, these claims remain statistical
in nature, being derived from assumptions about the transverse-velocity distributions. 

The degeneracy between mass and relative velocity can be lifted if precise astrometry is added to the photometry of the microlensing event.  
The size of the expected astrometric shift is small---of the order of milliarcseconds---but it is 
proportional to the angular Einstein radius $\theta_\mathrm{E}$, and therefore
if this small shift can be measured, then the mass of the lens can be determined unambiguously, as described in detail below. 

\subsection{Photometric Microlensing \label{subsec:photometric_microlensing} }

{\it Photometric\/} microlensing is the apparent transient brightening that results
as a background source passes almost directly behind a
foreground lens (see reviews by \citealt{Paczynski1996}, \citealt{Gaudi2012},
and \citealt{Tsapras2018}).  

With
$\vec\theta_\mathrm{S}$ and $\vec\theta_\mathrm{L}$ denoting the angular positions of the source and lens as seen by the observer, we can define a dimensionless source-lens separation
\begin{equation}
\vec u \equiv \frac{\vec\theta_\mathrm{S}-\vec\theta_\mathrm{L}}{\thetaE} \, .
\end{equation}

As the lens intervenes near the line of sight from the observer to the source,
the gravitational bending of light leads to a time-varying magnification
\begin{equation}
A(u) = \frac{u^2 + 2}{u \sqrt{u^2+4}} \,,
\label{eq:mu}
\end{equation}
which depends solely on $u = |\vec u|.$
This expression holds as long as the finite angular size of the source star can be neglected,
which we will assume in the following discussion. We will demonstrate in \S\ref{sec:finite} that
this is a valid approximation for the case analyzed in this paper.

If $F_\mathrm{S}$ is the intrinsic source flux, and $F_\mathrm{B}$ the background flux contributed by any other objects not resolved from the observed source star, the observed flux of the target for a specific telescope and filter is given by
\begin{equation}
  \label{eq:blend}
    F(t) = F_\mathrm{S} A[u(t)] + F_\mathrm{B} = F_\mathrm{base} \frac{A[u(t)] + g}{1+g}\,,
\end{equation}
where $F_\mathrm{base} \equiv F_\mathrm{S}+F_\mathrm{B}$ is the baseline flux and $g \equiv F_\mathrm{B}/F_\mathrm{S}$ is the specific blend ratio.

\subsection{Astrometric Microlensing \label{subsec:astrometric_microlensing} } 

Microlensing also produces an {\it astrometric shift\/} of the apparent position of the source. If we assume that we can observe the centroid of light formed by the images of the source without any contribution from other bodies such as the lens or other neighboring stars, its shift is described by the vector
\begin{equation}
\vec\delta(\vec u) = \frac{\vec u}{u^2+2}\,\theta_{\rm E} 
\label{eq:delta}
\end{equation}
\citep{Gould1992, Paczynski1998}.
In contrast to the photometric microlensing signature,
the astrometric signature is explicitly proportional to the angular Einstein
radius~$\thetaE$.

Moreover, while the magnification diverges (for a point-like source) as $u\to 0$,
the astrometric shift becomes maximal for $u=\sqrt{2}$. The light magnification falls rapidly with increasing $u$: for large separations, $u \gg 1$, the brightness enhancement, $A(u)-1$, 
falls as $1/u^4$. On the other hand, the centroid shift, given by Equation~(\ref{eq:delta}), decreases more slowly with $u$, and for large separations it falls only as $1/u$.  The astrometric perturbation thus has a considerably longer duration than the photometric signal.  For more details, see \citet{Dominik2000}, \citet{Sahu2014}, and \citet{Bramich2018}.

The photometric and astrometric signatures of a microlensing event are connected because they arise from the same source-lens trajectory, $\vec u(t)$. Specifically, 
if a fit to the light curve of a microlensing event already yields $u$ as a function of time, 
the astrometric data then provide a direct measurement of the angular Einstein radius $\theta_\mathrm{E}$, as well as the orientation angle of the trajectory.
\color{black}

\subsection{Parallax Effect and Proper Motion}
\label{sec:pardef}

For constructing the source-lens trajectory $\vec u(t)$, we have to consider the 
proper motions $\vec{\mu}_\mathrm{S}$ and $\vec{\mu}_\mathrm{L}$ of the source and lens objects,
as well as their parallaxes, $\pi_\mathrm{S}$ and $\pi_\mathrm{L}$.

Let $\vec{\gamma}(t)\,(1~\mbox{AU})$ denote the projection of the Earth's orbit onto a plane perpendicular to the direction toward the source star.
The apparent geocentric positions of source and lens star are then given by
\citep[cf.][]{An2002,Gould2004}
\begin{eqnarray}
\vec{\theta}_\mathrm{S}(t) & = & \vec{\theta}_{\mathrm{S},0} + (t-t_0)\,\vec{\mu}_\mathrm{S} - \pi_\mathrm{S}\,\vec{\gamma}(t) \nonumber\,, \\
\vec{\theta}_\mathrm{L}(t) & = & \vec{\theta}_{\mathrm{L},0} + (t-t_0)\,\vec{\mu}_\mathrm{L} - \pi_\mathrm{L}\,\vec{\gamma}(t) \,,
\end{eqnarray}
so that for $\vec{\theta}(t) \equiv \vec{\theta}_\mathrm{S}(t) - \vec{\theta}_\mathrm{L}(t)$, one finds
\begin{equation}
\vec{\theta}(t)  =  (\vec{\theta}_{\mathrm{S}} -  \vec{\theta}_{\mathrm{L}})_0 - (t-t_0)\,\vec{\mu}_\mathrm{LS} + \pi_\mathrm{LS}\,\vec{\gamma}(t)\,,
\end{equation}
where $\vec{\mu}_\mathrm{LS} \equiv \vec{\mu}_\mathrm{L} -\vec{\mu}_\mathrm{S}$ and $\pi_\mathrm{LS} \equiv \pi_\mathrm{L}-\pi_\mathrm{S}$ are the relative proper motion and relative parallax between lens and source, while $(\vec{\theta}_\mathrm{S}-\vec{\theta}_\mathrm{L})_0 \equiv \vec{\theta}_{\mathrm{S},0} - \vec{\theta}_{\mathrm{L},0}$.

Consequently, with the microlensing parallax parameter $\pi_\mathrm{E}=\piLS/\theta_\mathrm{E}$, $\vec{u}(t) \equiv \vec{\theta}(t)/\theta_\mathrm{E}$ takes the form
\citep[cf.\ ][]{Dominik2019}
\begin{equation}
\vec{u}(t) = \vec{u}_0 + (t-t_0)\,\dot{\vec{u}}_0 + \pi_\mathrm{E}\,\delta\vec{\gamma}(t) \, ,
\end{equation}
where
\begin{eqnarray}
 \vec{u}_0 \equiv  \vec{u}(t_0) & = &  \frac{(\vec{\theta}_\mathrm{S}-\vec{\theta}_\mathrm{L})_0}{\theta_\mathrm{E}} + \pi_\mathrm{E}\,\vec{\gamma}(t_0) \, , 
 \\
  \dot{\vec{u}}_0 \equiv  \dot{\vec{u}}(t_0)  & = & -\frac{\vec{\mu}_\mathrm{LS}}{\theta_\mathrm{E}} + \pi_\mathrm{E}\,\dot{\vec{\gamma}}(t_0) \, , \label{eq:dirdiff}
\end{eqnarray}
as well as 
\begin{equation}
  \delta\vec{\gamma}(t)  = \vec{\gamma}(t)-\vec{\gamma}(t_0)-(t-t_0)\,\dot{\vec{\gamma}}(t_0)\,.
  \label{eq:pargam}
\end{equation}
By construction, $\delta\vec{\gamma}(t_0) = 0$ and $\delta\dot{\vec{\gamma}}(t_0) = 0$.

By choosing $t_0$ so that $ \vec{u}_0 \perp \dot{\vec{u}}_0$, we can write $\vec{u}(t)$ in its components toward
northern and eastern directions as
\begin{eqnarray}
u_\mathrm{n}(t) & = &  \frac{t-t_0}{t_\mathrm{E}}\,\cos\psi - u_0\,\sin\psi + \pi_\mathrm{E}\,\delta\gamma_\mathrm{n}(t)\,, \nonumber \\
u_\mathrm{e}(t) & = &   \frac{t-t_0}{t_\mathrm{E}}\,\sin\psi + u_0\,\cos\psi + \pi_\mathrm{E}\,\delta\gamma_\mathrm{e}(t)\,,
\end{eqnarray}
where $u_0 \equiv |\vec{u}_0|$, $t_\mathrm{E} = 1/|\dot{\vec{u}}_0|$, and $\psi$ denotes the direction angle of $\dot{\vec{u}}_0$ measured from north toward east.
Alternatively, the trajectory can be parameterized as
\begin{eqnarray}
u_\mathrm{n}(t) & = &  \frac{t-t_0^\star}{t_\mathrm{E}^\star}\,\cos\psi^\star - u_0^\star\,\sin\psi^\star + \pi_\mathrm{E}\,\gamma_\mathrm{n}(t)\,, \nonumber \\
u_\mathrm{e}(t) & = &   \frac{t-t_0^\star}{t_\mathrm{E}^\star}\,\sin\psi^\star + u_0^\star\,\cos\psi^\star + \pi_\mathrm{E}\,\gamma_\mathrm{e}(t)\,,
\end{eqnarray}
where $u_0^\star \equiv |\vec{\theta}_\mathrm{S}(t_0^\star)-\vec{\theta}_\mathrm{L}(t_0^\star)|/\theta_\mathrm{E}$, $t_\mathrm{E}^\star = \theta_\mathrm{E}/|\vec{\mu}_\mathrm{LS}|$, and $\psi^\star$ denotes the direction angle of $-\vec{\mu}_\mathrm{LS} = \vec{\mu}_\mathrm{S}-\vec{\mu}_\mathrm{L}$ measured from north toward east.
Note that the starred quantities, $\psi^\star$, $\tE^\star$, $t_0^\star$ and $u_0^\star$, refer to parameters in the barycentric reference frame, whereas the corresponding
unstarred quantities refer to parameters as seen by an observer on Earth.

\begin{figure}
\begin{center}
\includegraphics[width=\columnwidth]{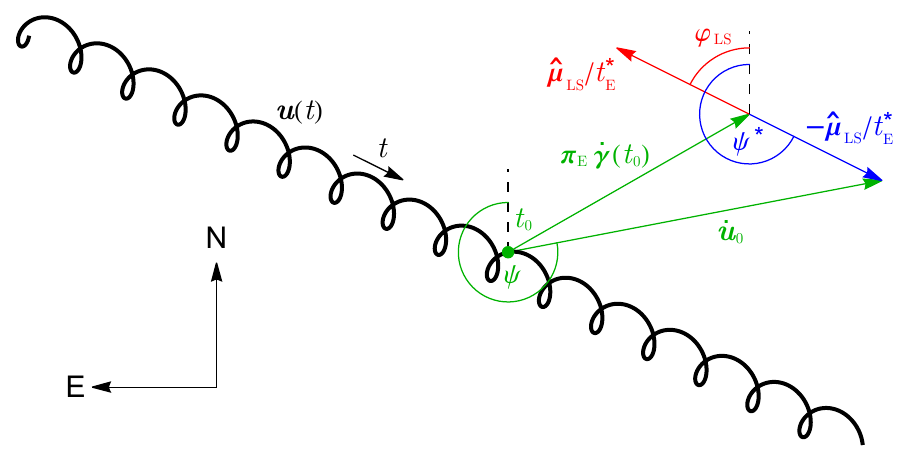}
\end{center}
\caption{Source-lens trajectory $\vec{u}(t)$ as seen by the observer, showing the effect of annual parallax. At epoch $t_0$, the tangent to the source-lens trajectory $\dot{\vec{u}}_0$ is not (anti-)parallel to the direction of the lens-source proper motion $\hat{\vec{\mu}}_\mathrm{LS}$, but differs by $\pi_\mathrm{E}\,\dot{\vec{\gamma}}(t_0)$ [Equation~(\ref{eq:dirdiff})], related to the orbital velocity of the Earth at $t_0$. Consequently, we distinguish the direction angles $\psi$, $\psi^\star$, and $\varphi_\mathrm{LS}$, referring to $\dot{\vec{u}}_0$, and $\mp \hat{\vec{\mu}}_\mathrm{LS}$, respectively. Furthermore, $t_\mathrm{E} = 1/|\dot{\vec{u}}_0|$
and $t_\mathrm{E}^\star = \theta_\mathrm{E}/|\vec{\mu}_\mathrm{LS}|$.
\label{fig:SLTgeo}
}
\end{figure}

This applies to any orthonormal reference frame; the only difference is in the specific angle, e.g., $\psi_\mathrm{eq}$, $\psi_\mathrm{ecl}$, $\psi_\mathrm{gal}$ (and $\psi^\star_\mathrm{eq}$, $\psi^\star_\mathrm{ecl}$, $\psi^\star_\mathrm{gal}$) for equatorial, ecliptic, and galactic coordinates, respectively, which are related by a rotation of the coordinate axes at the target position.

While most discussions of photometric microlensing events choose an {\em ecliptic\/} coordinate frame, the observed astrometric data are more easily described in an {\em equatorial\/} coordinate frame, and therefore we adopt the latter in the following analysis. In this frame,
\begin{equation}
    \varphi_{\rm LS} = \psi^\star_\mathrm{eq}+ 180^\circ
\end{equation}
gives the position angle of the proper motion of the lens with respect to the source $\vec{\mu}_\mathrm{LS}$, measured from equatorial north toward east.  We illustrate the geometry of the source-lens trajectory $\vec{u}(t)$ in Figure~\ref{fig:SLTgeo}.

\subsection{Measuring the Lens Mass}

As stated above, the only useful physical parameter in a typical microlensing event is the
timescale $t_\mathrm{E}^\star = \theta_\mathrm{E}/\mu_\mathrm{LS}$, which is 
the time
it takes the source to traverse the radius of the Einstein ring,
which itself depends on the lens mass, $M_\mathrm{lens}$, and the lens-source parallax, $\pi_\mathrm{LS}$. 

However, for long-duration events, the annual parallax tends to lead to prominent
departures in the photometric signature \citep{Gould1992, Alcock1995}, so that a
microlensing parallax parameter $\piE \equiv \pi_\mathrm{LS}/\thetaE$ can be inferred.
Coincidently, the BH-mass lenses tend to imply such long-duration events.
On the other hand, the astrometric signature is proportional to the angular Einstein radius
$\thetaE$, so that by combining  photometric and astrometric observations, $M_\mathrm{lens}$, $\pi_\mathrm{LS}$, and $\vec \mu_\mathrm{LS}$ become fully decoupled.
Specifically, with $\piE$ from the photometry and $\thetaE$ from the astrometry, the definition of $\thetaE$, Equation~(\ref{eq:theta}), immediately gives us

\begin{equation}
\Mlens = \frac{\theta_\mathrm{E}}{\pi_\mathrm{E}} \frac{c^2 (\rm1 \, AU)}{4G} \ = \frac{\theta_\mathrm{E}}{\kappa \pi_\mathrm{E}} \, , 
\label{eq:mass}
\end{equation}
\noindent where $\kappa = 4 G / [c^2 ({\rm1 \, AU})] \simeq 8.144 \, {\rm mas}\,M_\odot^{-1}$.

In the case of microlensing toward the Galactic bulge, the source often lies at the distance of the bulge itself, which can be verified from its baseline position in a color-magnitude diagram (CMD)\null. If spectroscopic observations are available in addition to baseline photometry---as is the case for the event discussed in this paper---a more accurate source distance can be determined.
The lens-source relative parallax, $\pi_{LS}=\pi_\mathrm{E} \, \theta_\mathrm{E}$, 
can then be used to estimate the distance to the lens, using Equation (\ref{eq:distances}). As a bonus, the event timescale gives a direct measure of the relative transverse velocity of lens and source (which, for stellar remnants, might include ``kicks'' received in SN explosions).  This method thus provides independent measurements of three separate physical parameters of the lens: its mass, distance, and transverse velocity.

\subsection{Characteristics of Astrometric Deflections}

\begin{figure}[b]
\includegraphics[width=0.47\textwidth]{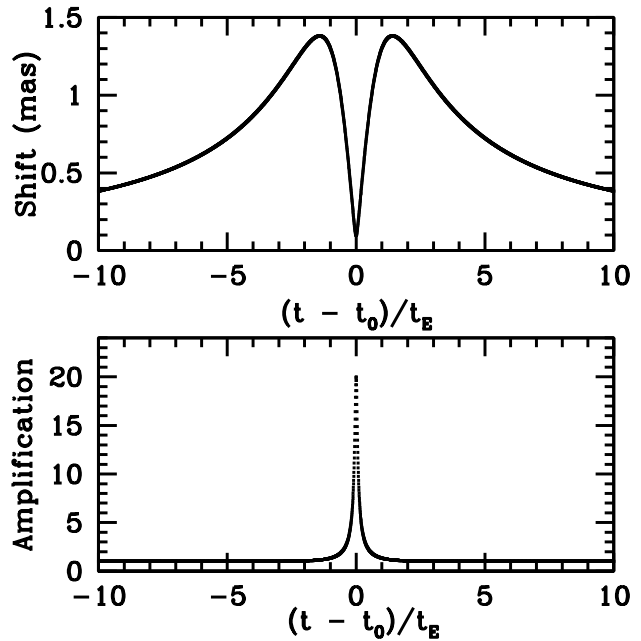}
\caption{
Astrometric shift (top panel) and light {\amplification} (bottom panel) for a microlensing event produced by a $5\,M_\odot$ black hole at a distance of 2~kpc passing in front of a background star at 8~kpc. The assumed minimum impact parameter is $u_0=0.05$. 
The maximum astrometric shift of the source is $\sim$1.4~mas, at $u=\sqrt2$, and the maximum {\amplification} is $\sim$20, at the time of closest angular approach. Note the much longer duration for the astrometric shift, compared to that of the light {\amplification}. 
\label{fig:typicalevent}
}
\end{figure}

Some features of astrometric deflections under various scenarios are described by \citet{Dominik2000}. To illustrate a typical case, we show in Figure~\ref{fig:typicalevent} the calculated astrometric shifts and light {\amplification} for a nominal event of a BH lens of mass $5\,M_\odot$, at a distance of 2~kpc from the Sun, passing in front of a background source situated in the Galactic bulge at a distance of 8~kpc. The closest angular approach is assumed to be at a separation of $0.05~\thetaE$. In this case, the size of the angular Einstein ring is $\theta_\mathrm{E} \simeq 4$~mas, so that the  maximum astrometric shift is $\sim$1.4~mas (occurring at a separation of $u=\sqrt{2}$), and the maximum light {\amplification} is a factor of $\sim$20 (at closest angular approach). 

As Figure~\ref{fig:typicalevent} illustrates, and as discussed in \S\ref{subsec:astrometric_microlensing}, the duration of the astrometric deflection is considerably longer than that of the  photometric {\amplification}. This makes it necessary to carry out the astrometric measurements over a longer time interval than the photometry.  Although the deflection measured at any given epoch provides in principle an estimate of $\theta_\mathrm{E}$, it is necessary to observe at multiple epochs in order to separate the shifts caused by microlensing from those caused by the proper motion of the source; observations at a late epoch are particularly useful for this purpose.  The figure also shows that the astrometric shift is close to zero at the time of highest {\amplification}; therefore observations near the photometric peak are also very useful to constrain the source proper motion.  

\begin{figure}[h]
\includegraphics[width=0.47\textwidth]{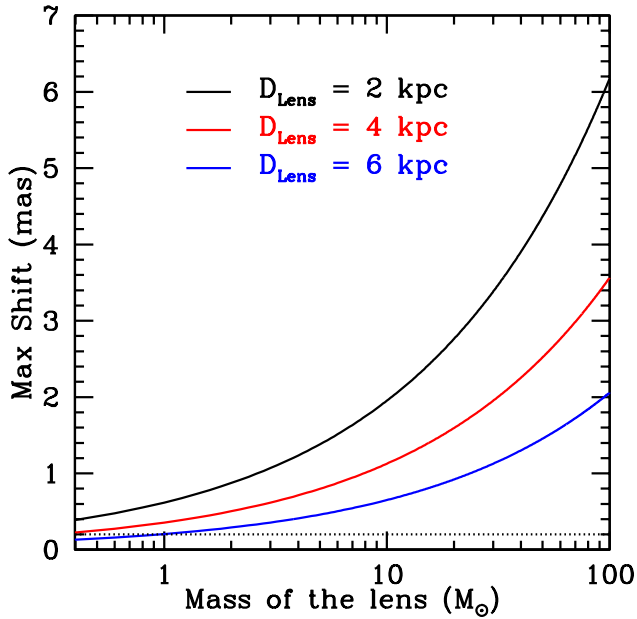}
\caption{Maximum astrometric shift of a source in the Galactic bulge at 8~kpc as a function of lens mass, for lens distances of 2, 4, and 6~kpc. Since the maximum astrometric shift occurs at a lens-source separation of $u=\sqrt2$, while the maximum light {\amplification} occurs at the minimum value of $u$, all high-{\amplification} microlensing events will pass through the point of maximum astrometric deflection. High-S/N imaging with \HST\/ allows measurements of astrometric shifts with a precision of $\sim$0.2~mas per observation epoch, shown by the dotted line at the bottom.
\label{fig:bulgeevents}
}
\end{figure}

In Figure~\ref{fig:bulgeevents}, we plot the maximum astrometric shifts for a source in the Galactic bulge at 8~kpc, as functions of lens mass. The lenses are assumed to be located at distances of 2 and 4~kpc (``disk'' lenses), and 6~kpc (``bulge'' lenses).  The dotted line at the bottom shows the nominal astrometric precision of 0.2~mas achievable with high-SNR \HST\/ imaging, as discussed below.  Therefore the deflection is {\it detectable\/} at 1$\sigma$ per epoch for lens masses down to $\sim\!0.5\,M_\odot$, except at lens distances larger than 6~kpc. The most favorable situation for a precise mass measurement, of course, would be for a nearby, high-mass lens.

\subsection{High-Precision Astrometry}

Although an unambiguous determination of lens mass is possible from a combination of photometry and astrometry, as we have just discussed, the expected astrometric shifts are extremely  small, of the order of milliarcseconds or less. \HST\/ has demonstrated its capability to carry out sub-milliarcsecond astrometry through a variety of techniques. For example, high-S/N \HST\/ observations of isolated sources were used to achieve sub-milliarcsecond accuracy, leading to the measurement of proper motions for several distant hypervelocity stars \citep{Brown2015}. A collection of sources was used as probes to achieve an astrometric accuracy of $\sim$12 microarcseconds, in order to measure the transverse velocity of M31 \citep{Sohn2012}.  Spatial-scan techniques have been used to achieve an astrometric accuracy of  $\sim$30 microarcseconds \citep{Casertano2016, Riess2018} in the trigonometric parallax of Cepheids used for accurate determination of $H_0$, and to measure the distance to the globular cluster NGC\,6397 \citep{Brown2018}. Recently, our group used the astrometric-microlensing technique to measure the mass of
the nearby white dwarf Stein\,2051\,B, achieving an astrometric precision of $\sim$0.2~mas per epoch \citep{Sahu2017}.  \citet{Kains2017} looked for astrometric deflections in \HST\/ observations of 10 microlensing events with timescales of $<$50 days. They achieved an astrometric precision of 0.2~mas per epoch (but did not detect any deflections). From the ground, \citet{zurlo2018} used VLT to measure the
mass of Proxima Centauri through astrometric microlensing.
\citet{Lu2016} employed the Keck telescope to  look specifically for isolated BHs by monitoring three microlensing events, where they achieved a final positional error of 0.26 to 0.68~mas. The timescales of those events were 60 to 160 days, and there were no detections of astrometric deflections.

\section{In Search of Isolated Black Holes with \HST}

\subsection{Astrometry of Long-Duration Microlensing Events \label{subsec:our_search} }

In 2009, we
began a multi-cycle \HST\/ program of astrometry of long-duration microlensing events in the direction of the Galactic bulge in order to
detect isolated BHs and measure their masses. 
Our aim is to select events having timescales $\gtrsim$200~days, light curves showing no evidence for a light contribution by a luminous lens, and preferably a high {\amplification} factor. We then obtain high-resolution \HST\/ imaging as the events proceed, in order to measure the astrometric deflections of the background sources. To date we have monitored eight long-duration events. For some of them, there is no clear detection of an astrometric signal, but our data analysis is still in progress, and the results will be discussed in separate publications.  In the present paper we analyze and discuss our findings for an event that clearly shows a large astrometric deflection, consistent with a high-mass lens. 

\subsection{MOA-2011-BLG-191\slash  OGLE-2011-BLG-0462}

MOA-2011-BLG-191\slash  OGLE-2011-BLG-0462 (hereafter designated \OB) was a long-duration and high-{\amplification} microlensing event in the direction of the Galactic bulge. It was discovered independently by both MOA and OGLE ground-based microlensing survey programs, and announced by both teams  nearly simultaneously on 2011 June 2, through their public-alert websites.\footnote{MOA alerts: \url{https://www.massey.ac.nz/~iabond/moa/alerts}. OGLE alerts: \url{http://ogle.astrouw.edu.pl/ogle4/ews/2011/ews.html}} The target was also covered by the Wise Microlensing Survey. Table~\ref{table:basicdata} gives details of this remarkable event. 

\begin{deluxetable*}{lcc}[bt]
\tablecaption{Basic Data for \OB\ Microlensing Event \label{table:basicdata} }
\tablehead{
\colhead{Parameter} &
\colhead{Value} &
\colhead{Sources \& Notes\tablenotemark{a}}
}
\startdata
Event designation (MOA)              & MOA-2011-BLG-191   & (1) \\
Event designation (OGLE)             & OGLE-2011-BLG-0462  & (1) \\
J2000 right ascension, $\alpha$      & 17:51:40.2082  & (2) \\
J2000 declination, $\delta$          & $-29$:53:26.502 & (2) \\
Galactic coordinates, $(l,b)$        & $359\fdg86, -1\fdg62$ & (2) \\
Baseline F606W magnitude             &  $21.946\pm0.014$        & (3) \\
Baseline F814W magnitude             &  $19.581\pm0.012$        & (3) \\
Baseline ($\rm F606W-F814W$) color   &  $ 2.365\pm0.026$       & (3) \\
Peak magnification, $A_{\rm max}$    & 369 & (4) \\
Date of peak magnification, $t_0$    & 2011 July 20.825 & (4) \\
Timescale, $t_\mathrm{E}$            & $270.7\pm11.2$ days & (4) \\
\enddata
\tablenotetext{a}{Sources and notes: (1) MOA and OGLE websites;
the event was first alerted by MOA; (2) This paper, from astrometric analysis in 
\S\ref{subsec:astrometricanalysis} in \Gaia\/ EDR3 frame at average epoch 2013.5; (3) This paper, Vegamag scale, from photometric analysis in \S\ref{subsec:photometricanalysis}; (4) This paper, from Table \ref{table:lensproperties}.} 
\end{deluxetable*}

\OB\ occurred in an extremely crowded Galactic bulge field, less than $2^\circ$ from the Galactic center. 
The observed peak {\amplification} factor of this event was only about 20 in the ground-based data, but this was strongly diluted by blending with neighboring stars. It soon became apparent, based on findings disseminated through internal communications in the microlensing groups, that the undiluted event actually had an extremely high {\amplification} factor, approaching 400. Blending also made the apparent timescale of the event appear shorter than the actual value, which was inferred to be longer than 200 days. 
It was clear from the ground-based observations that there was blending, for two reasons.
First, the light curve for a typical event has a characteristic shape that is completely determined by the timescale and the maximum {\amplification}, except for distortions due, e.g., to the lens-source relative parallax. The shape of the observed light curve was inconsistent with the expected shape unless the light at baseline was highly diluted by a blend, thus implying that the real {\amplification} was much larger than the observed value.  Second, as the source brightened, its centroid position in the ground-based images was seen to change, again consistent with blending with a neighboring star. Note that this shift is due simply to blending and scales with the separation of the two stars; it is unrelated to the much smaller relativistic deflection of the source itself, which is discussed below.

\begin{figure}
\centering
\includegraphics[width=0.47\textwidth]{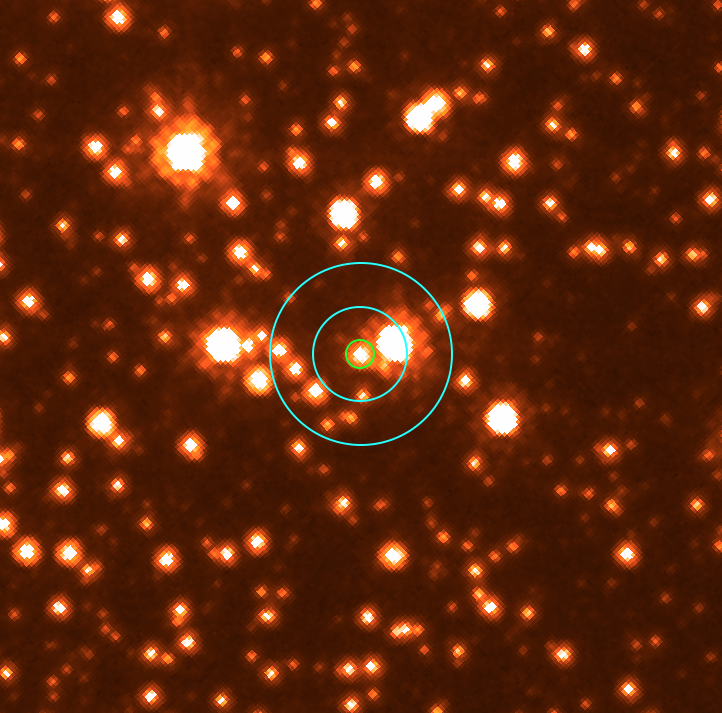}
 \caption{\HST\/ image in the F814W ($I$-band) filter of an $8'' \times 8''$ region centered on \OB, obtained at our final epoch in 2017 August. North is at the top, east on the left. Encircled in green is the source star, now returned to baseline luminosity. The site is resolved into the source, a much brighter neighboring star $0\farcs4$ to the WNW, and several nearby fainter stars. The inner cyan circle has a diameter of $1''$, corresponding to the typical best seeing in ground-based microlensing survey images; the outer cyan circle's diameter is $2''$, which is not unusual seeing. The source, bright neighbor, and several fainter stars are generally blended in ground-based frames, and the blending increases with seeing. 
\label{fig:8x8zoomin}
}  
\end{figure}

Figure~\ref{fig:8x8zoomin} shows an $8''\times8''$ region centered on the source, as imaged by us in the F814W ($I$-band) filter by \HST\/ with its Wide Field Camera~3 (WFC3)\null. The source star is encircled in green. A conspicuous neighbor, nearly 20 times brighter than the un\amplified{} source, lies at a separation of only $0\farcs4$. The cyan circles in the figure have diameters of $1''$ and $2''$, corresponding to the generally best seeing in the ground-based survey observations, and more typical seeing, respectively. Thus, in ground-based images, the source is indeed blended with the bright neighbor and a number of fainter stars, depending on the seeing.

High-{\amplification} events are generally very sensitive to perturbations due to planets around the lensing objects \citep{Mao1991,Griest1998}. Thus considerable interest was aroused by \OB\ among groups engaged in searches for such planets.  As a result, intensive photometric monitoring of this event was carried out by multiple groups, providing valuable data for our analysis.

\section{\HST\/ Observations }

The \OB\ event satisfies all the selection criteria for our \HST\/ follow-up program described in \S\ref{subsec:our_search}, and thus we triggered our observing sequence. 
Our project had a ``non-disruptive'' target-of-opportunity status, requiring a lead time of about two to three weeks from activation to the first observations. The first-epoch \HST\/ data were obtained on 2011 August~8, some 19~days after the peak light {\amplification} on 2011 July~20. The {\amplification} was still reasonably high ($\sim$12, corresponding to $u\simeq 0.08$), so that the expected
astrometric deflection was $\delta \simeq 0.04\,\theta_{\rm E} $
[see Equation~(\ref{eq:delta})], i.e., close to zero at this epoch, but its correct value is taken into account in the model described in \S\ref{sec:fullmodeling}.
Subsequent \HST\/ observations indicated departure from a linear proper motion for the source. Thus we continued the imaging, ultimately over an interval of over six years, long enough for robust separation of the relativistic deflection from proper motion\null. Table~\ref{table:journal} gives the \HST\/ observing log.

\begin{deluxetable*}{clccccc}
\tablecaption{Journal of \HST\/ Wide Field Camera 3 Observations \label{table:journal} }
\tablehead{
\colhead{Epoch}           &
\colhead{Date} &
\colhead{MJD}  & 
\colhead{Year}  &
\colhead{Proposal} &
\colhead{No.\ Frames}  & 
\colhead{No.\ Frames}   \\
\colhead{} &
\colhead{}         & 
\colhead{}         & 
\colhead{}         &
\colhead{ID} & 
\colhead{in F606W\tablenotemark{a}}    &
\colhead{in F814W\tablenotemark{a}}          
}
\startdata
1 & 2011 Aug 8  & 55781.7 & 2011.600 & GO-12322 & 4 & 5  \\
2 & 2011 Oct 31 & 55865.2 & 2011.829 & GO-12670 & 3 & 4   \\
3 & 2012 Sep 9  & 56179.2 & 2012.689 & GO-12670 & 3 & 4   \\
4 & 2012 Sep 25 & 56195.3 & 2012.733 & GO-12986 & 3 & 4   \\
5 & 2013 May 13 & 56425.8 & 2013.364 & GO-12986 & 3 & 4   \\
6 & 2013 Oct 22 & 56587.2 & 2013.806 & GO-13458 & 3 & 4   \\
7 & 2014 Oct 26 & 56956.1 & 2014.816 & GO-13458 & 3 & 4  \\
8 & 2017 Aug 29 & 57994.7 & 2017.660 & GO-14783 & 3 & 4   \\
\enddata
\tablenotetext{a}{Individual exposure times ranged from a minimum of 60~s at Epoch~1, to a maximum of 285~s at later epochs.}
\end{deluxetable*}

All our \HST\/ observations were obtained with the UVIS channel of WFC3, whose CCD detectors provide a plate scale of $\rm39.6 \, mas \, pixel^{-1}$. To avoid buffer dumps during the orbital visibility period and thus maximize observing efficiency, we used the UVIS2-2K2C-SUB subarray, giving a field of view (FOV) of $80''\times80''$. This FOV is large enough to provide dozens of nearby astrometric reference stars surrounding the primary target. 

The WFC3 detectors are subject to an increasing amount of degradation of their charge-transfer efficiency (CTE) as they are exposed to the space environment. The chosen subarray aperture places the target in the middle of the left half of the UVIS2 CCD, which lessens the impact of imperfect CTE relative to a  placement closer to the center of the FOV\null.  Nevertheless, a time-dependent correction for CTE must still be applied in the astrometric analysis of the images.

Our \HST\/ observations were taken at a total of eight epochs, strategically scheduled for measurement and characterization of the astrometric deflections. At each epoch, we obtained images in two filters (to verify the achromatic nature of the event, and to test for blending by very close companions): ``$V$'' (F606W) and ``$I$'' (F814W). At the initial epoch, when the source was bright, we obtained nine exposures, four in F606W and five in F814W\null. At each subsequent epoch, using longer integration times because of the fading of the source, we obtained seven exposures, three in F606W and four in F814W\null.  Individual exposure times were adjusted to take into account the brightness of the source and the orbital visibility of \HST, and ranged from a minimum of 60~s at the first epoch to a maximum of 285~s at the later epochs. The telescope pointing was dithered by $\sim$200 pixels ($\sim$$8''$) between individual exposures; this allowed retention of a common set of reference stars in all the exposures, in order to mitigate errors in the distortion solution. To maximize the S/N for the most crucial astrometric measurements, we separated Epochs~3 and 4 by only 16~days in 2012 September, around the time when the deflection was expected to be near maximum.

\begin{figure*}
\centering
\includegraphics[width=\textwidth]{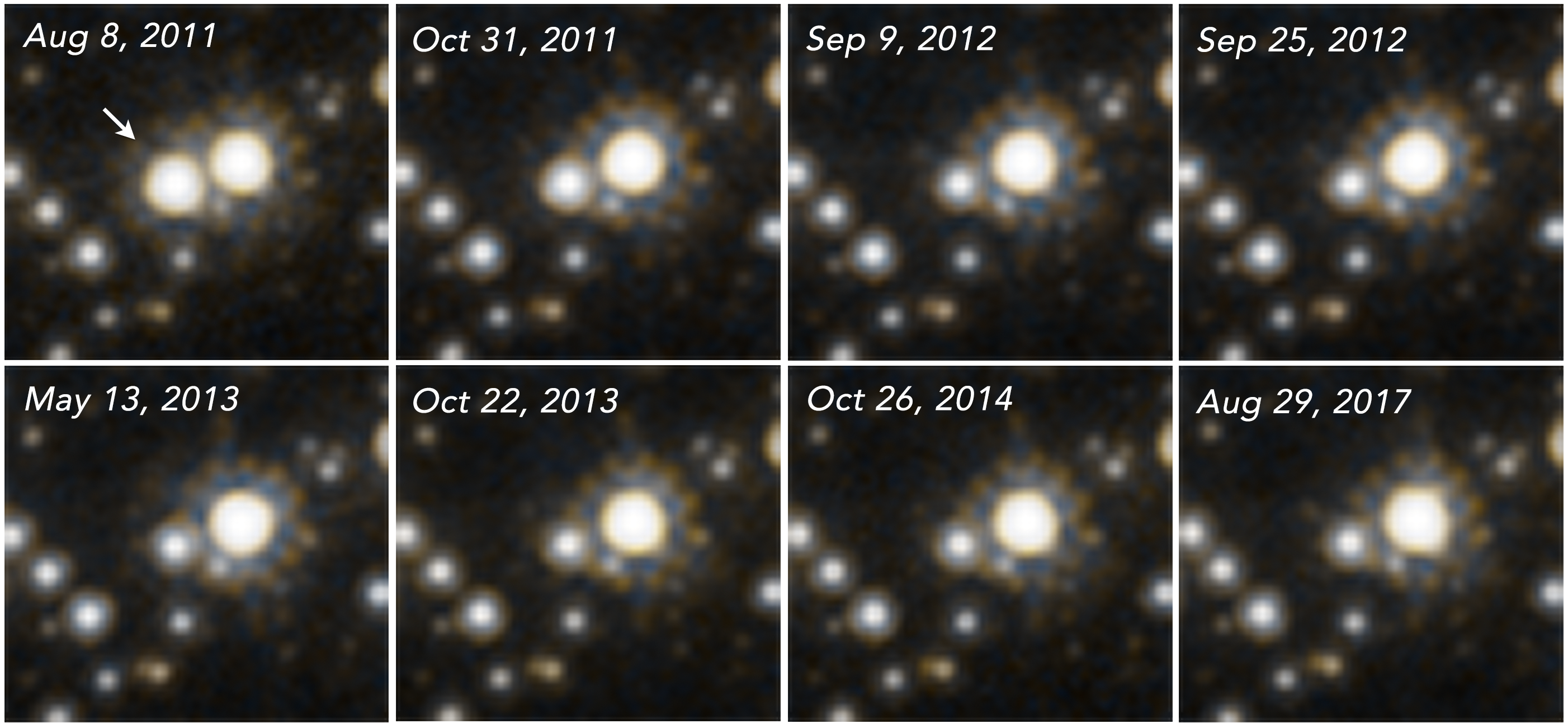}
\caption{
$2\farcs1 \times 2\farcs0$ cutouts around \OB\ as observed by \HST\/ at all 8 epochs. 
Exposures were taken in F606W and F814W from 2011 to 2017; see Table~\ref{table:journal} for details. The source star is marked by an arrow in the first-epoch (top left) image. At this epoch, on 2011 August 8, the magnification was a factor of $\sim$12. The maximum astrometric deflection occurred at Epochs~3 and 4, when the photometric magnification was only about 10\%.  In 2017, the source had returned very close to its un\amplified{} brightness and undeflected position. 
\label{fig:postagestamps}
} 
\end{figure*}

Figure~\ref{fig:postagestamps} zooms in on the field around the source in Figure~\ref{fig:8x8zoomin}, showing a $2\farcs1 \times 2\farcs0$ region as observed at all eight epochs. The bright source is marked with an arrow in the Epoch~1 (top left) image, and it can be seen to fade in the subsequent frames. The astrometric deflection was highest at Epochs~3 and 4, even though the photometric {\amplification} was only about 10\% at this epoch.  There was very little photometric change in the subsequent epochs,  but the astrometric deflections remained detectable until Epoch~7, demonstrating the need for astrometric monitoring over a much longer duration than the photometric-variability period.

\section{\HST\/ Data Analysis }

\subsection{Image Processing}

We used the flat-fielded and CTE-corrected ({\tt\_flc}) images produced by the Space Telescope Science Institute pipeline reductions \citep{Sahu2021, Dressel2021} for the analysis.  As noted above, WFC3 suffered from increasingly poor CTE during this period, so it was essential to take it into account. The {{\tt\_flc}} products were produced using the v2.0 pixel-based CTE model described by \citet{Anderson2021}.

\subsection{Astrometric Analysis \label{subsec:astrometricanalysis} }

To measure stellar positions in individual frames, we used an updated version of the star-measuring algorithm described in \citet{Anderson2006}.  The routine goes through each exposure pixel by pixel, and identifies as a potential star any local maximum that is sufficiently bright and isolated.  The routine uses the spatially variable effective point-spread functions (PSFs) provided at the WFC3/UVIS website{\footnote{\url{https://www.stsci.edu/hst/instrumentation/wfc3/data-analysis/psf}}} to fit the PSF to the star images in the individual {\tt \_flc} exposures, in order to determine a position and flux for each star in the raw pixel frame of that exposure.  Finally, the positions are corrected for geometric distortion using the distortion solutions provided by
\citet{Bellini2011}.

\begin{figure}
\epsscale{1.20}
\plotone{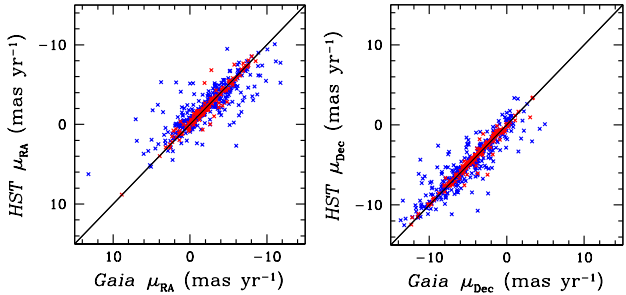}
\vskip -4cm
\caption{Proper motions of the reference stars used in our analysis derived from our \HST\/ observations with a 6-year baseline, versus the \Gaia\/ proper motions.
Red points represent stars brighter than $G=18$, for which the \Gaia\/ errors are typically $<$$0.2\, \masyr$, and blue points represent fainter stars for which the errors are larger.
Our analysis uses the \Gaia\/ reference frame, so it is natural that there is good agreement between them.  But the individual \HST\/ measurements have higher 
precision, particularly at fainter magnitudes. 
\label{fig:pm_ref_stars}
}
\end{figure}

As the positions of individual stars are expected to change during the $\sim$6-year course of our observations due to their proper motions, we needed to determine their proper motions to properly specify the reference frame.  For this, we began with the \Gaia\/ Early Data Release~3 (EDR3) \citep{Gaia2021} positions and motions for the bright but unsaturated \HST\/ stars in the field.  The reference frame was constructed to place the bright star close to \OB\ at the center of the reference frame at $(x,y)=(1000,1000)$ at the 2016.0 epoch, with a plate scale of $40\,\rm mas\, pixel^{-1}$ and north up.  (Note that the \Gaia\/ catalog could be incomplete in this region because of the high source density.)  Using the \Gaia\/ positions and motions, we determined the position for each \Gaia\/ star in this frame at each epoch in order to properly transform the distortion-corrected observations at that epoch into the reference frame.  This ensures that the proper motions that we derive represent absolute proper motions.  After this initial set-up of the reference frame based on the brighter stars, we incorporated high-precision \HST\/ stars that were too faint to be found with high precision in the \Gaia\/ catalog and solved for their accurate positions and motions. We then used their time-dependent positions to improve the reference frame.  Even after allowing individual solutions to improve on the basis of \HST\/ observations, there remains very good agreement between our proper motions and those of \Gaia.  Figure~\ref{fig:pm_ref_stars} plots the proper motions of our reference stars derived from our \HST\/ observations against the \Gaia\/ proper motions, where the red points are for brighter stars with $G < 18$, and blue points are for fainter stars with $G \ge 18$. The agreement is imperfect, of course, since the \HST\/ observations have a 6-year baseline and have higher S/N in individual measurements, resulting in higher accuracy in proper-motion measurements, particularly at fainter magnitudes.  The agreement is better for the brighter 
sample, for which \Gaia\/ proper-motion errors are typically smaller ($<$$0.2\, \masyr$). Note that the
 \OB\ source itself is too faint at baseline for inclusion in the \Gaia\/ catalog. 

\begin{figure*}
\plotone{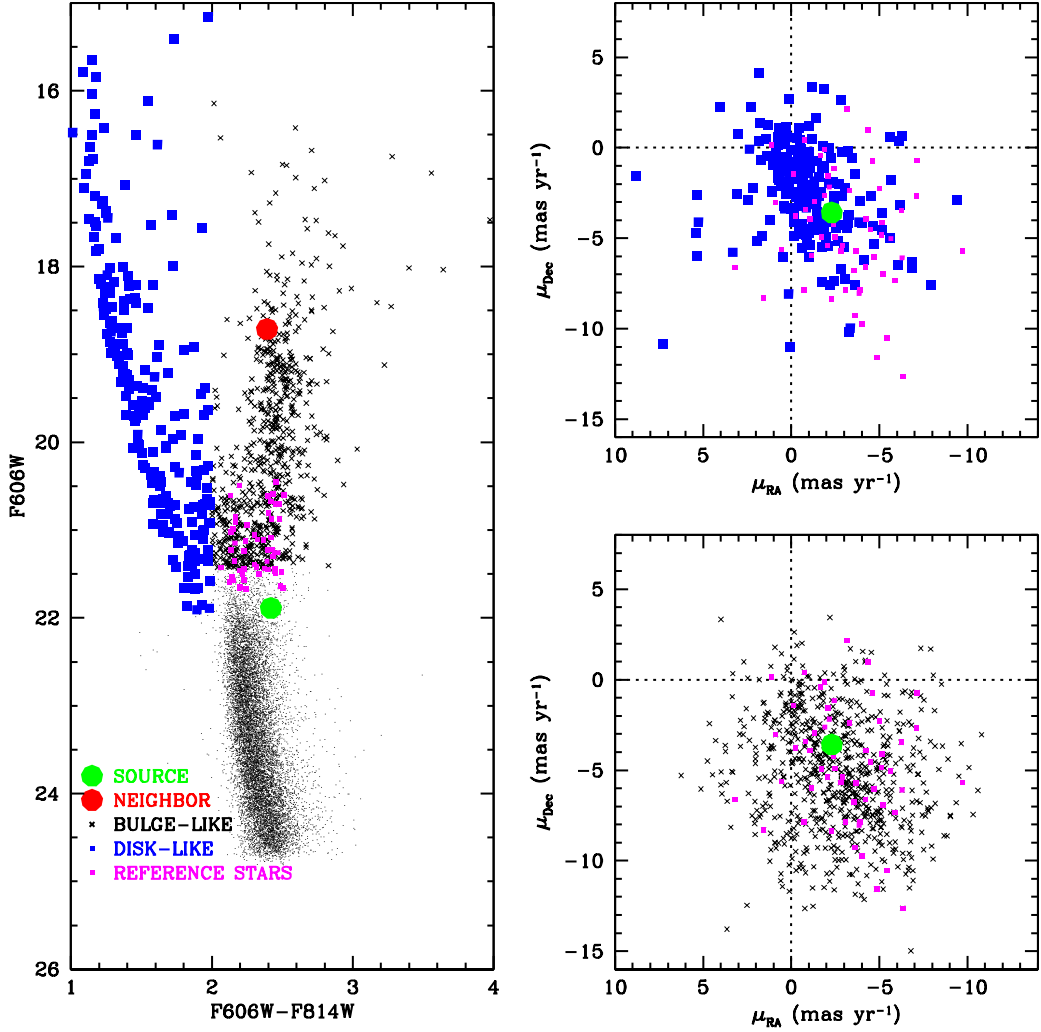}
\caption{Left panel: color-magnitude diagram [$m_{\rm F606W}$ versus $(m_{\rm F606W}-m_{\rm F814W})$] for all stars in the \HST/WFC3 field. The main-sequence turnoff occurs at $m_{\rm F606W}\simeq 22$, above which the disk and the bulge split into two sequences:  the redder stars are mainly bulge objects, while the bluer ones are mainly disk objects. Stars marked by blue squares are selected as ``disk-like'' stars, and the stars marked by black crosses are selected as ``bulge-like.'' 
The position of the (unmagnified) microlensed source is shown as a green dot.  
The nearby bright star $0\farcs4$  away and about 20 times brighter than the unmagnified source is shown as a red dot.
Magenta points are the astrometric reference stars used in our analysis. 
Top right panel: proper motions of the disk-like stars (blue), the reference stars (magenta), and the source (green dot). Bottom right panel: proper motions of the bulge-like stars (black crosses), the reference stars (magenta), and the source (green dot). Only bulge-like stars (shown in magenta) with brightness similar to the observed brightness of the star were used in the final astrometric transformations.  This ensures that the reference stars are very similar to the target source star, and uncorrected CTE and small parallax effects should cancel in the differential astrometric measurements.
\label{fig:CMD}
}
\end{figure*}

In the next step of our analysis, we used only the \HST\/ observations, because the \Gaia\/ measurements have much higher uncertainties for the fainter stars, and the \HST\/ observations have a longer baseline of 6 years compared to the 3~years of \Gaia. In this step of the transformation, we used stars (1)~with brightness similar to the average brightness of the target, (2)~with color similar to the source's color (see Figure~\ref{fig:CMD}), and (3)~lying within 350 pixels of the source. The first criterion minimizes any residual shift caused by CTE effects. We note that we already used the most recent CTE correction software for our analysis. Since the CTE effects on the position measurements are differential, using linear transformations based on stars of similar brightness should remove any residual CTE effects. (It is worth noting here that the images with the highest astrometric deflection were taken when WFC3/UIVS was young, and when CTE losses were small.) The second criterion ensures that the stars used in the transformation belong to the bulge, which helps in minimizing errors due to parallax, as described in more detail below.  The third criterion minimizes residuals in the distortion solution. 

We employed an iterative procedure to measure the positions and proper motions of the stars, starting from the revised values in each iteration.
We rejected the highest-sigma point after each complete iteration. We repeated this procedure until the highest-sigma point was no more than a preset tolerance, for which we adopted $6\sigma$.  Only a small number of points were rejected by this procedure, mostly affected by cosmic-ray hits on the detector. Then at each epoch the reference-star positions were corrected for proper motion, and the positions of the source were determined relative to this adjusted frame. The estimated uncertainty in the position of the source star relative to the adjusted frame is $\sim$0.4 mas in each individual exposure. 
 
\begin{figure*}
\includegraphics[width=\textwidth]{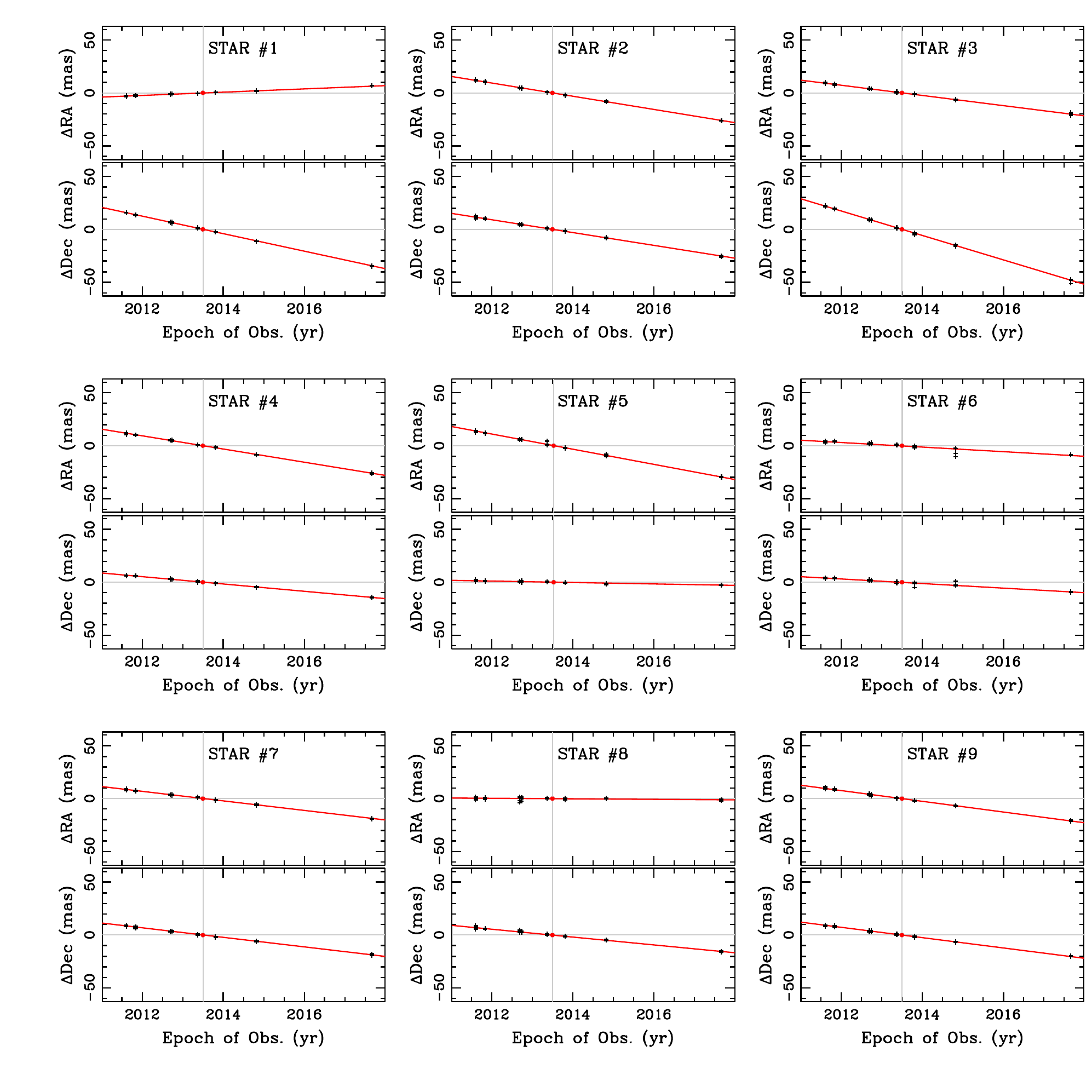}
\caption{
Motions of nine representative astrometric reference stars in right ascension and declination. The red lines are linear fits to the proper motions of the stars.
\label{fig:pm_ref}
}
\end{figure*}

\begin{figure}
\epsscale{1.20}
\plotone{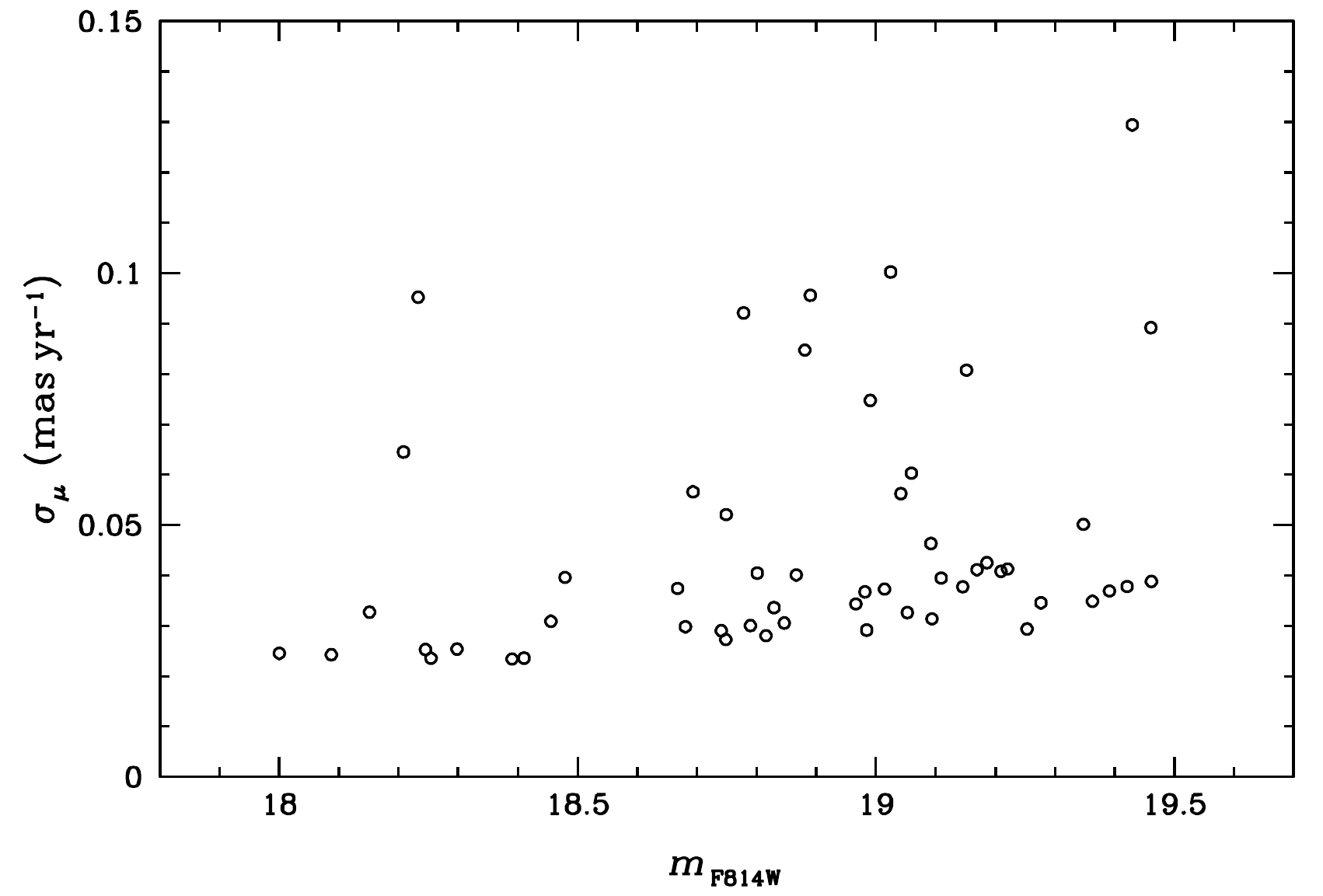}
\caption{
Proper-motion errors of the reference stars.  The $y$-axis shows the proper-motion error, defined as  
$\sigma_\mu = [\sigma_{\mu x}^2 + \sigma_{\mu y}^2]^{0.5}$,
where $\sigma_{\mu x}$   and $\sigma_{\mu y}$ are the proper-motion errors along the $x$ and $y$ axes, which are parallel to right ascension and declination, respectively.  Reference stars cover a range of 1.5 mag around the (unamplified) magnitude of the source; their proper motion uncertainties vary significantly from object to object, with a modest systematic increase with magnitude.  
\label{fig:pm_error}
}
\end{figure}

\begin{figure}
\epsscale{1.10}
\plotone{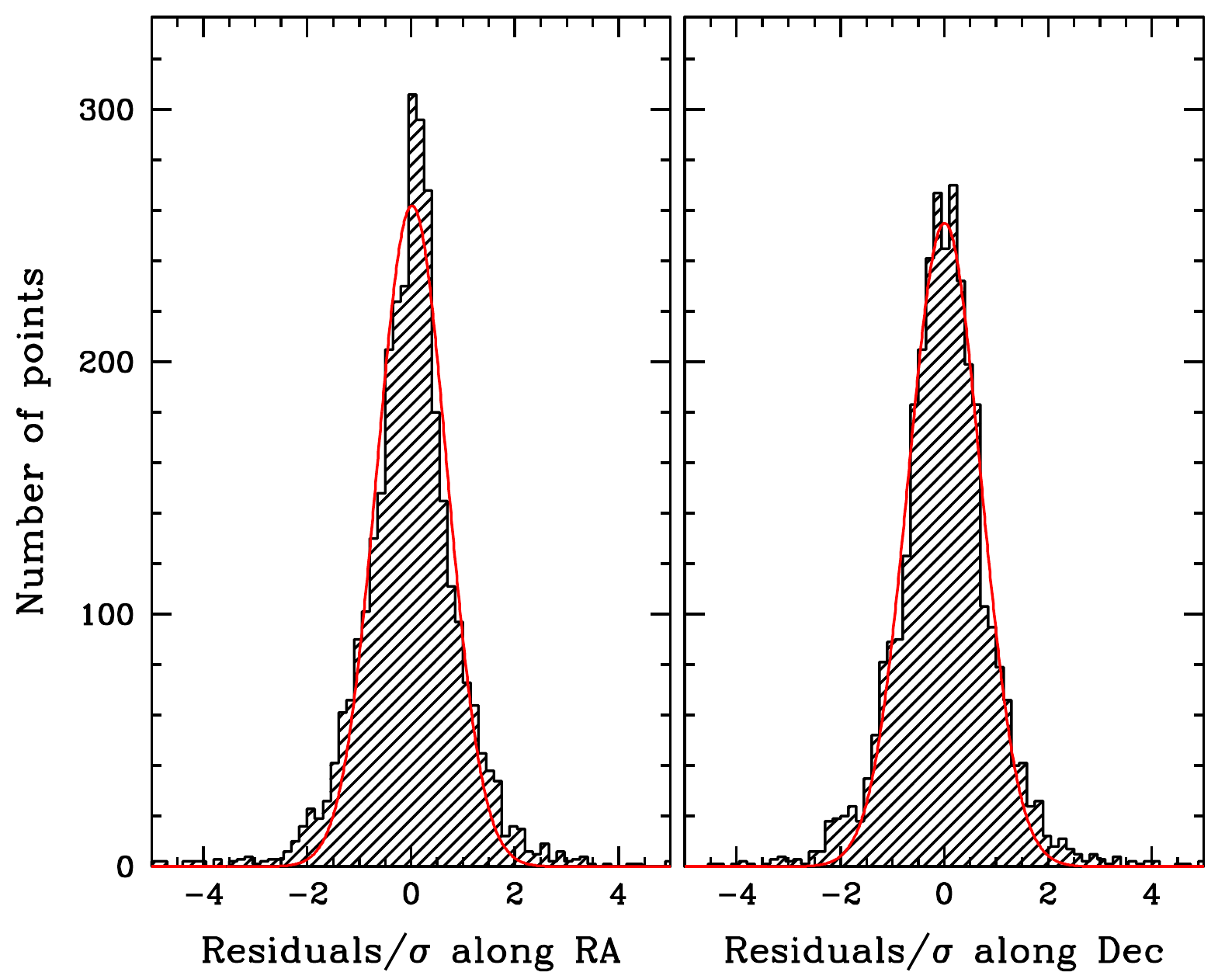}
\caption{Histogram of the position residuals for all the reference stars in RA and Dec from their proper motion solution, as measured in each image separately, and scaled with the measured dispersion for that star.  The scaled residuals closely follow a standard Normal distribution (shown by the red curves).  As noted in the caption to Figure~\ref{fig:pm_error}, reference stars cover a range of magnitudes around the unamplified source magnitude.  
\label{fig:pm_res_hist}}
\end{figure}

As an illustration, Figure~\ref{fig:pm_ref} shows the proper motions as measured for nine representative stars. Figure~\ref{fig:pm_error} shows the errors in the proper motion measurements of the reference stars. We note that, all the reference stars are within about 1.5~mag of each other (See Figure~\ref{fig:CMD}).  Figure~\ref{fig:pm_res_hist} shows the the histograms of the residuals of each measurement from that star's proper motion solution along the RA and Dec directions.  Both distributions are consistent with a Gaussian distribution (the red curve).   As shown by previous similar studies, the final reference-frame positions are expected to be internally accurate to better than 0.01 pixel \citep{Anderson2008,Bellini2015}. 

We specifically solved for the proper motions of the reference stars, but we ignored their parallaxes. The reason for adopting this approach is the following. 

Our choice of reference stars ensures that a large fraction of them belong to the Galactic bulge, and hence have similar parallactic motion as the source star. We note that the source parallax is small to begin with ($\lesssim$0.2 mas). Then the source position is referenced to stars chosen to be at comparable distance, so any remaining impact of the source parallax on the astrometry or photometry of the event is expected to be negligible. 

As described earlier, there is a bright star $\sim$10 pixels away from our source.  The target star is close to the brightness of this neighbor in the first two epochs and slowly fades to its nominal brightness, at which point it is about 3~mag fainter than the bright neighbor.  For accurate astrometry of the source, we wanted to make sure that the position measurements of the source are not affected by the presence of this bright neighbor. So we wanted to subtract the PSF of the bright star before measuring the positions of the source at every epoch. However, subtracting the bright star is not just a matter of subtracting a standard PSF\null. The separation is $\sim$10 pixels and the available library PSFs go out to 12 pixels, and are tapered and not very accurate in the wings. Thus we needed to make a more extended PSF model.

\begin{figure}
\begin{center}
\includegraphics[height=2.0in]{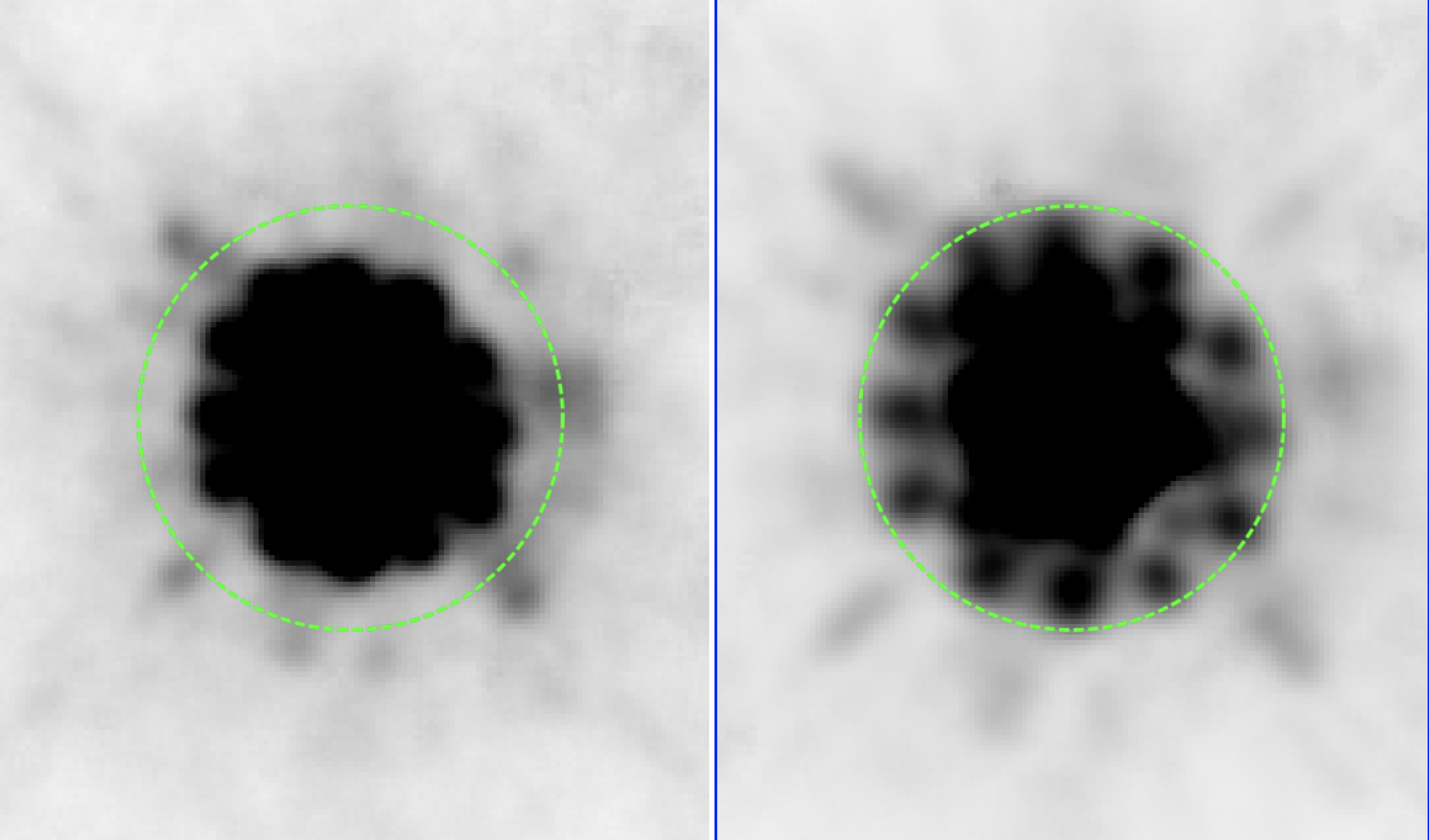}
\end{center}
 \caption{
The stacked F606W (left) and F814W (right) PSFs, with the 10-real-pixel radius shown in green. Note that the F606W PSF has a diffraction spike bump very close to the location of the source star.  The bump for the other diffraction spikes does not show up in the residual images shown in the second row of Figure~\ref{fig:psf}, which implies that it is subtracted well beneath the source, making the position measurements of the source more robust. The F814W PSF has a very strong radial gradient (along with azimuthal structure) at the location of the source.  Without subtracting a high-fidelity model PSF, there would be some impact on the measurement of the source positions.  And as the source moves relative to the neighbor, then the source would move across the PSF halo, which could introduce artificial shifts if the PSF of the bright neighbor is not subtracted well.
 \label{fig:stacked_psfs}
 }
\end{figure}

To make an extended PSF, we carefully selected stars that (1) are within $\sim$350 pixels of the bright neighbor, (2) have brightness and color similar to the neighbor, and (3) are fairly isolated.  We found 18 such stars (excluding the bright star itself), which provided a good sample to make the required extended PSFs. We used the images of these 18 stars to produce a separate well-sampled, extended PSF for each individual exposure. For illustration, Figure~\ref{fig:stacked_psfs} shows the stacked PSFs in F606W and F814W\null. We have taken particular care to make sure that the PSF is well characterized in the wings since the source lies in the wings of the bright star, and subtracting the wings correctly is crucial for accurate astrometry.

\begin{figure*}
\begin{center}
\includegraphics[height=3.5in]{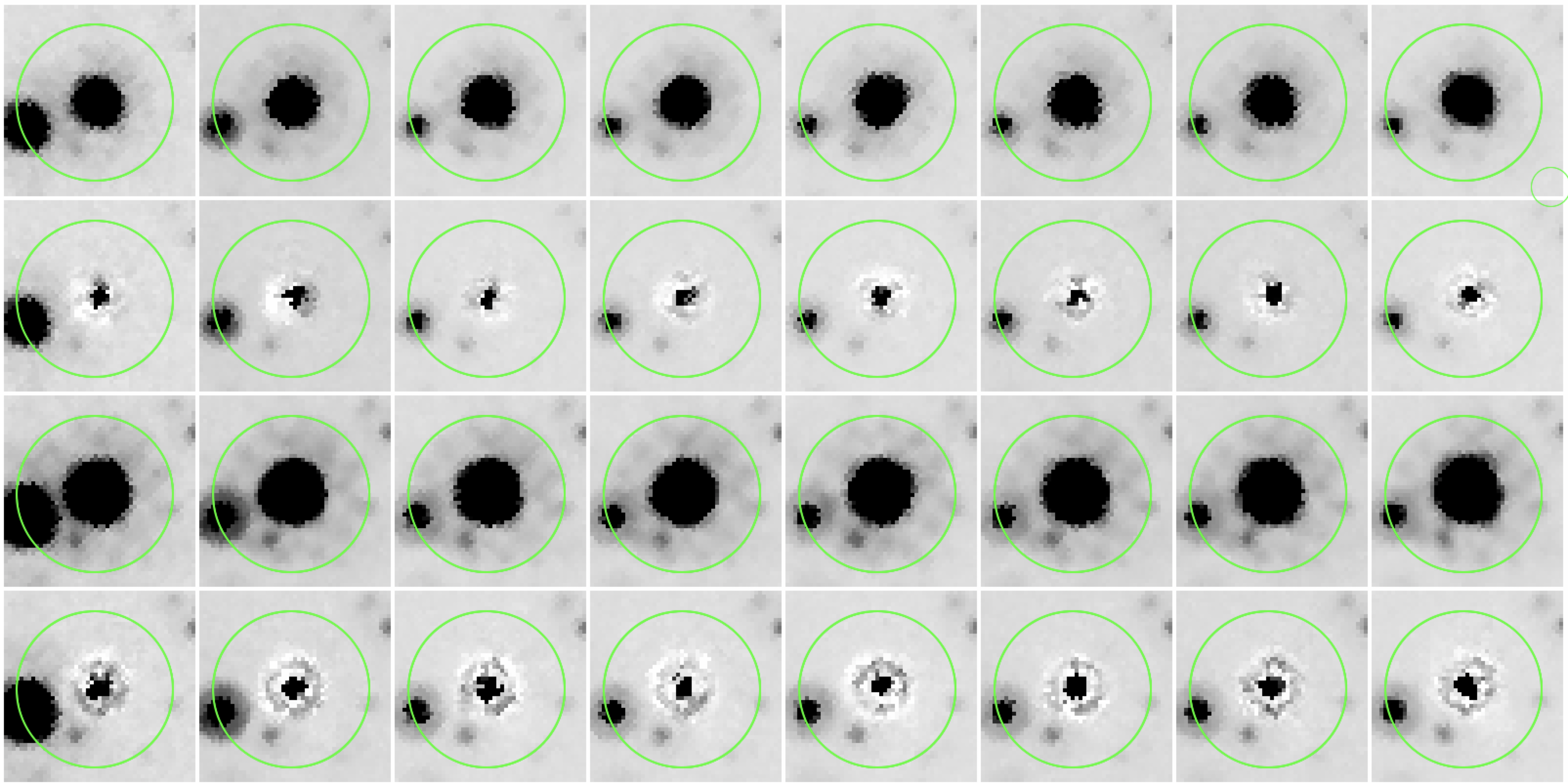}
\end{center}
\caption{The source star lies in the PSF wings of a bright neighbor, marked in these frames with green circles with a radius of 10~pixels. Since the available WFC3 library PSFs do not extend to this large a radius,  a special PSF extending to 20 pixels was constructed using the same \HST\/ images. This PSF was used to subtract the neighbor before measuring the position of the source in each exposure. The top row shows the original stacked F606W images of Epochs~1 to 8 (from left to right), and the second row shows the PSF-subtracted images. The third row shows the original stacked F814W images, and the bottom row shows the subtracted images. The stacks are overbinned by a factor of 2 to show details. Care was taken to assure that the PSF is well characterized in the wings of the neighbor star where the source lies. 
\label{fig:psf}}
\end{figure*}

We then took this PSF model for each exposure and subtracted it from the neighbor star in each exposure. Figure~\ref{fig:psf} shows the original images (first and third rows) and the subtracted images (second and fourth rows) in the F606W and F814W filters. The residuals are very small, particularly in the wings of the PSF\null.  We found that the astrometric position of the source changes by $\sim$0.03 pixel (1.2~mas) after this subtraction, which could have a significant effect on the mass determination of the lens; so this extra step of neighbor subtraction was crucial in improving the analysis/results. The resulting astrometric positions of the source were used for further analysis as described in the next section.

\subsection{Photometric Analysis \label{subsec:photometricanalysis} } 

In addition to the measured positions, the analysis algorithm provides PSF-based photometry of all the stars in the field. To set a calibrated zero-point, we used standard aperture photometry to determine the fluxes of a few isolated stars in the field within an aperture with a 10-pixel radius.  These fluxes were then corrected to an infinite aperture, using encircled-energy measurements from \citet{Calamida2021}, and the photometric zero-point in the image headers ({\tt PHOTFLAM}) was used to convert these fluxes to the Vegamag scale. The mean difference between these values and the values obtained by the PSF fitting was then applied to all of the PSF magnitudes to convert them to Vegamag. 

\begin{figure}
\includegraphics[width=0.47\textwidth]{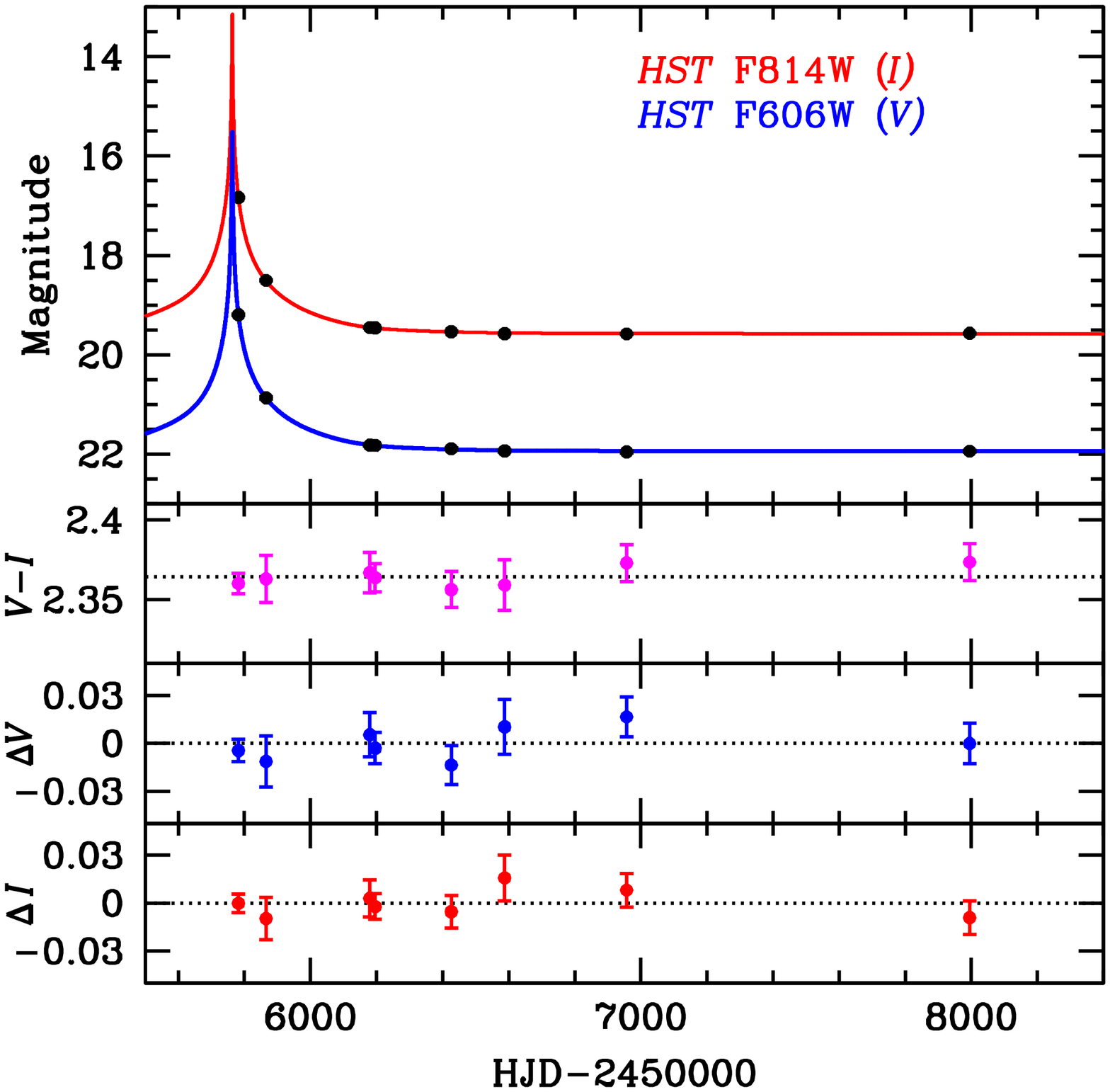}
\caption{Top panel: Photometry of \OB, obtained with \HST\/ over 6 years at 8 epochs in the F606W ($V$) and F814W ($I$) filters (black filled circles), along with our final model fits from \S8.2 (red and blue curves). Errors on the photometry are smaller than the plotting points. 
The second panel plots the observed values of $V-I$, showing that the source color remained constant to
within $\sim$0.01~mag during the entire 6-year duration. 
The third and fourth panels show that the residuals, $V_{\rm obs}-V_{\rm model}$ and
$I_{\rm obs}-I_{\rm model}$, at the epochs of the \HST\/ observations
are consistent with zero within the measurement uncertainties
(see \S\ref{subsub:second_approach} for more details and discussion).}
\label{fig:HST_lightcurve}
\end{figure}

As described above, the source star lies on the wings of the PSF of the neighboring bright star. So, for accurate photometry of the source, it was critical to correctly subtract the light contribution from the bright neighbor. The photometry for the source was carried out after subtracting the superposed flux from the neighbor star using a high-fidelity PSF as described above. The resultant time-series \HST\/ photometry of the source is shown in Figure~\ref{fig:HST_lightcurve}, along with the model light curve described below in \S\ref{sec:fullmodeling}.  There is no detectable color change as the event progresses and the star fades: the color of the source has remained constant to within 0.01 mag during the 6 years of observations with \HST\null. There is also no detectable blending as described in more detail in \S\ref{subsub:second_approach}. The source at baseline brightness has apparent magnitudes of $m_{\rm F606W} = 21.946 \pm 0.014$ and $m_{\rm F814W} = 19.581 \pm 0.012$.

We made a stack of all the images for each filter for every epoch. 
The \HST\/ images allow us to detect and measure magnitudes of stars as faint as $V\simeq25$. Figure~\ref{fig:CMD} shows the CMD based on this photometry, where we also show the position of the source and the bright neighbor $0\farcs4$ away.

\section{Ground-based Light Curve}

\OB\ was monitored photometrically by several ground-based observatories. The coverage by MOA, OGLE, and Wise Microlensing Survey extended over several years. Moreover, as a high-magnification event, it attracted intensive monitoring by a number of additional ground-based telescopes---especially around the time of peak brightness, where the microlensing light curve is sensitive to planet detection. Table~\ref{table:photometry_journal} gives a journal of the photometric observations and data used in our analysis.
We use the data re-reduced by the surveys and other groups
\citep{Udalski1992, Bond2001, Sackett2004, Gould2006, Tsarpas2009, Dominik2010}.

\begin{figure*}
\begin{center}
\includegraphics[width=\textwidth]{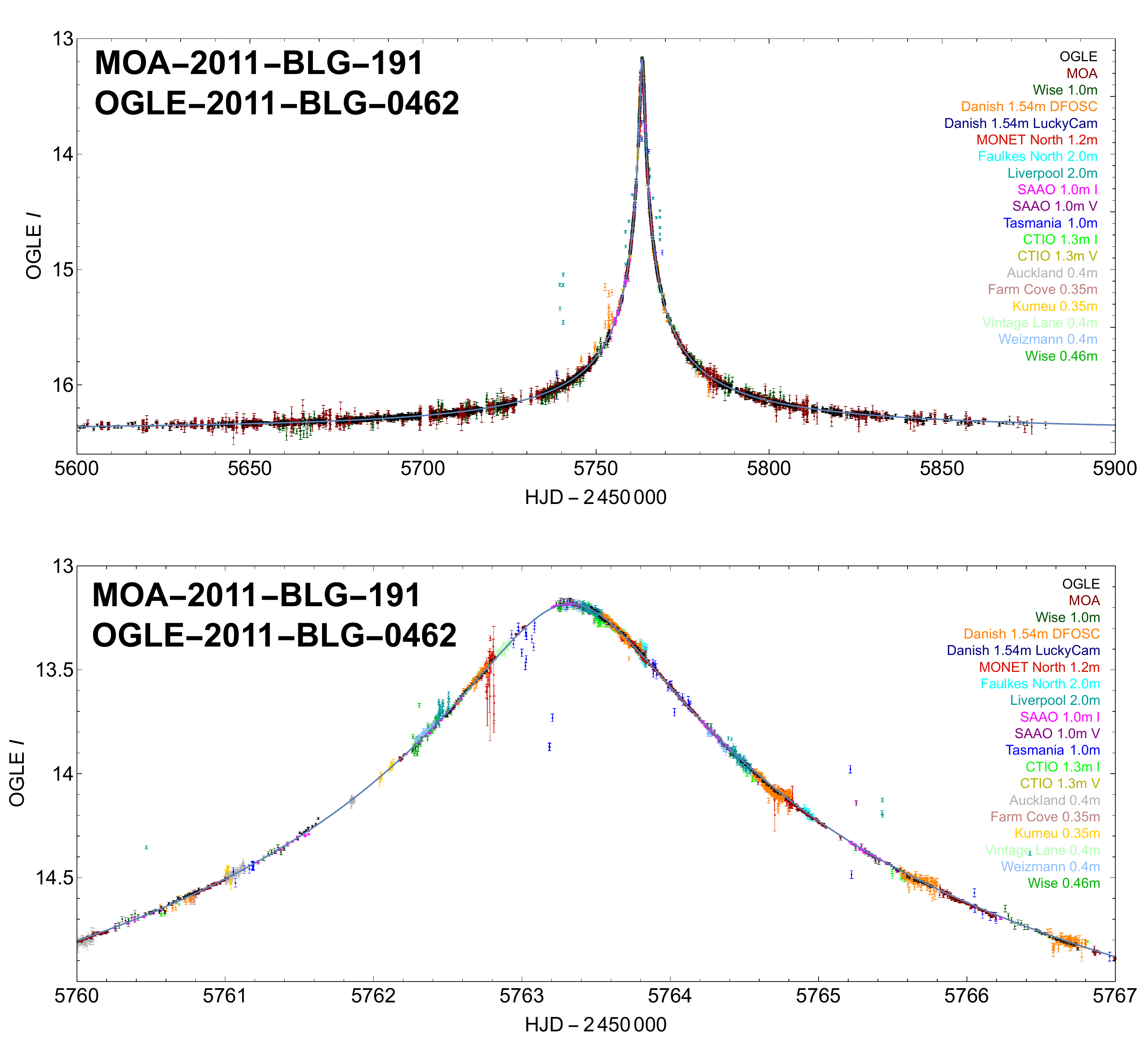}
\end{center}
\caption{Ground-based photometric observations of \OB, along with a best-fitting model light curve, shown over a 300-day interval in the upper panel and over a zoomed-in region covering seven days around peak magnification in the bottom panel. All data have been transformed to OGLE $I$ magnitudes according to the inferred baseline magnitudes and blend ratio from the common model.
\label{fig:AllPhotDataPlot}
}
\end{figure*}

Figure \ref{fig:AllPhotDataPlot} shows the light curve, both over a 300-day interval (top panel), and zooming in on the seven days around peak magnification (bottom panel). Superposed is our model fit to the light curve, from the analysis described in the next two sections.

\begin{deluxetable*}{lccCccllc}[bt]
\tablecaption{Journal of Ground-Based Photometry of \OB\ \label{table:photometry_journal} }
\tablehead{
\colhead{Data Set\hfill } &
\colhead{Telescope} &
\colhead{Aperture} &
\colhead{Filter}&
\colhead{No.\ of}&
\colhead{Date Range}&\\
&
\colhead{Location}&
\colhead{[m]}&&
\colhead{Observations}&
\colhead{($\mathrm{HJD}-2450000$)}
}
\startdata
MOA		& New Zealand & 1.8& R&	46040 & $3824.093 \ldots 9441.117$ \\
OGLE		&Chile& 	1.3 & I & 15546 & $5260.855 \ldots 8787.509$ \\
Wise Survey & Israel & 1.0 & $I$ & 953 & $5658.529 \ldots 5722.543$ \\
Danish DFOSC &Chile& 1.54 	& I & 921 & $5744.804 \ldots 5782.554$ \\
Danish LuckyCam & Chile & 1.54 	& {\rm broad} &   10 & $5738.624 \ldots 5742.706$\\
MONET North & Texas, USA & 1.2 & I	 &  214  & $5762.722 \ldots 5764.824$\\
Faulkes North &Hawaii, USA& 2.0 & {\rm SDSS}\ i' &	  99 & $5763.777 \ldots 5768.955$\\
Liverpool & Canary Islands, Spain& 2.0  & {\rm SDSS}\ i' &  254 & $5739.521 \ldots 5768.438$\\
SAAO 1.0m & South Africa & 1.0 & I &	  611 & $5751.264 \ldots 5777.315$\\
SAAO 1.0m & South Africa &1.0 & V &	   44 & $5758.454 \ldots 5765.385$\\
U.\ Tasmania& Australia& 1.0 & I&	    60  & $5761.055 \ldots 5769.110$\\
CTIO & Chile& 1.3 & I &		  226 & $5757.511 \ldots 5772.785$ \\
CTIO & Chile& 1.3& V &		   27 & $5757.515 \ldots 5763.807$ \\
Auckland &New Zealand& 0.4	& R	&   160 & $5759.873 \ldots 5778.893$  \\
Farm Cove &New Zealand & 0.35 & {\rm unfiltered} &	    37 & $5741.856 \ldots 5761.856$  \\
Kumeu Obs. & New Zealand & 0.35 & R	&    63 &$5759.803 \ldots 5762.134$ \\
Vintage Lane & New Zealand & 0.4 & {\rm unfiltered} &	    60 & $5762.792 \ldots 5767.941$  \\
Weizmann & Israel & 0.4 & I & 167 & $5762.281 \ldots 5764.423$\\
Wise & Israel & 0.46 & I & 142 & $5762.266 \ldots 5763.456$ \\
\enddata
\tablecomments{The data were acquired by the MOA \citep{Bond2001}, OGLE \citep{Udalski2015},
Wise Microlensing Survey \citep{Shvartzvald2016}, MiNDSTEp \citep{Dominik2010}, RoboNet \citep{Tsarpas2009}, PLANET \citep{Sackett2004}, and $\mu$FUN \citep{Gould2006} teams.}
\end{deluxetable*}

\section{Blending, Relative Parallax, and Lens Trajectory \label{sec:blending} }

The combination of ground-based photometric monitoring with long-term astrometric and photometric measurements from \HST\/ affords the ability to constrain all aspects of this event and obtain high-quality measurements of its parameters. In this section, we describe some characteristics of the data, illustrating the key information that can be obtained from photometry and astrometry separately through heuristic considerations.  

\subsection{Photometric Blending}
\label{subsub:blending}

Ground-based photometry of \OB\ suffers from significant amounts of blending with neighboring stars, as shown in our \HST\/ images  (Figure~\ref{fig:8x8zoomin}). Moreover, the amount of blending changes markedly depending on image quality. There is a bright neighbor star only $0\farcs4$ away, along with two fainter stars within $0\farcs5$ of the source. At larger separations, there are three more stars within $1''$, whose combined brightness is greater than that of the baseline source, and there are several more stars within $1\farcs5$ which are also brighter than the source.  Since in the available ground-based imaging data the measured image quality is seldom better than $ 1'' $, the ground-based photometry will always include the light from at least the three closest stars within $ 0\farcs5 $. In a fraction of the data, taken under poorer seeing conditions, the source is blended with an increasing number of neighbors.  

Our \HST\/ images allow us to place constraints on the expected blending parameter, $ g $, defined as the ratio of the flux from neighbors included in the photometry to the flux from the unmagnified source itself [Equation~(\ref{eq:blend})], in the ground-based observations. The bright neighbor is 18.88 times brighter than the source at baseline in F814W, and the contribution from the two fainter stars is an additional 0.19 times that of the source.  Thus the expected blending factor due to these three stars is $g=19.07$ in the F814W filter. Since the bright neighbor is similar in color to the source (see Figure~\ref{fig:CMD}), we adopt this value of $g$ for both OGLE ($I$~band) and MOA ($R$~band) data as an initial estimate, but keep it as a variable in our analysis.  The final values (\S\ref{subsec:astrometricanalysis}) differ significantly between OGLE and MOA, possibly due in part to differences in processing between the two data sets.

The effect of blending can be reduced substantially by basing the photometry on difference images.  It can be further reduced by restricting the analysis to images taken under good seeing.  It is obvious, however, that variable blending will affect the noise characteristics of ground-based photometry; it is difficult to include such effects deterministically because of the imperfect knowlegde of the blending (at the sub-percent level) for individual images.  

\subsection {Heuristic Considerations}\label{subsec:heuristic}

\subsubsection{Photometric Constraints on Parallax and Lens Trajectory \label{subsubsec:photometricconstraints} }

As discussed in \S\ref{subsec:astrometric_microlensing}, the light curve of a long-duration microlensing event such as \OB\ can show distortion by the relative parallactic motions of the source and lens \citep[e.g.,][]{Gould1992, Alcock1995}.  Specifically, the light curve is sensitive to $\piE$ and $\phil$, because their combination modifies the relative path of source and lens, and thus the shape of the light curve.  We note that, in our formalism, {\phil} corresponds to the position angle (PA) of the path of the lens relative to the source without parallax in equatorial 
coordinates (not to be confused with the instantaneous path of the lens at the time of closest angular approach in ecliptic coordinates; see \S\ref{subsec:photometric_microlensing}).

In principle, a sufficiently accurate light curve can provide good constraints on both $ \pi_{\mathrm E} $ and $ \phil$.  However, as discussed in the previous subsection, the photometry is significantly affected by blending. We attempted to model the light curve alone, but found that it can be fitted with a range of parameter combinations, in which the values of $ \pi_{\mathrm E} $ and $ \phil$ are strongly correlated.  In addition, the derived value of $\pi_{\mathrm E} $ varies with the specific subset of photometric data chosen for analysis, as well as with the assumed blending factor for those data. The derived value of $\piE$ ranges from 0.07 to 0.12, with larger values corresponding to larger values of $ \phil$, ranging from $330\degr$ to $358\degr$.  The reason is that increasing  \phil\ makes the lens move in a more northerly direction as seen in the
bottom panel of Figure~\ref{fig:lens_path}. Since parallax is predominantly in the east-west direction, in order to produce a fixed change in $u$, the value of $\pi_\mathrm{E}$ has to increase with \phil, so that the change in position due to parallax can compensate for a more northerly motion of the lens. Several different combinations of these quantities can reproduce the observed light curve, with differences between solutions of the order of 1~mmag at early and late times, and $\sim$5~mmag near the peak.  Systematic differences in the data at this level could be caused by small variations in blending associated with changes in the ground-based seeing, or other minor secular variations in the photometry.  Therefore we conclude that, when photometry alone is used to constrain the parameters of the event, only a reliable joint constraint on $ \pi_{\mathrm E} $ and $ \phil $ can be derived.  Fortunately, astrometry provides a robust independent estimate of $\phil$, allowing us to break this degeneracy and determine the two quantities separately. 
\subsubsection{Astrometric Deflection and Orientation of the Relative Motion}
\label{subsubsec:astrometricconstraints}

\begin{figure}
\epsscale{1.1}
\plotone{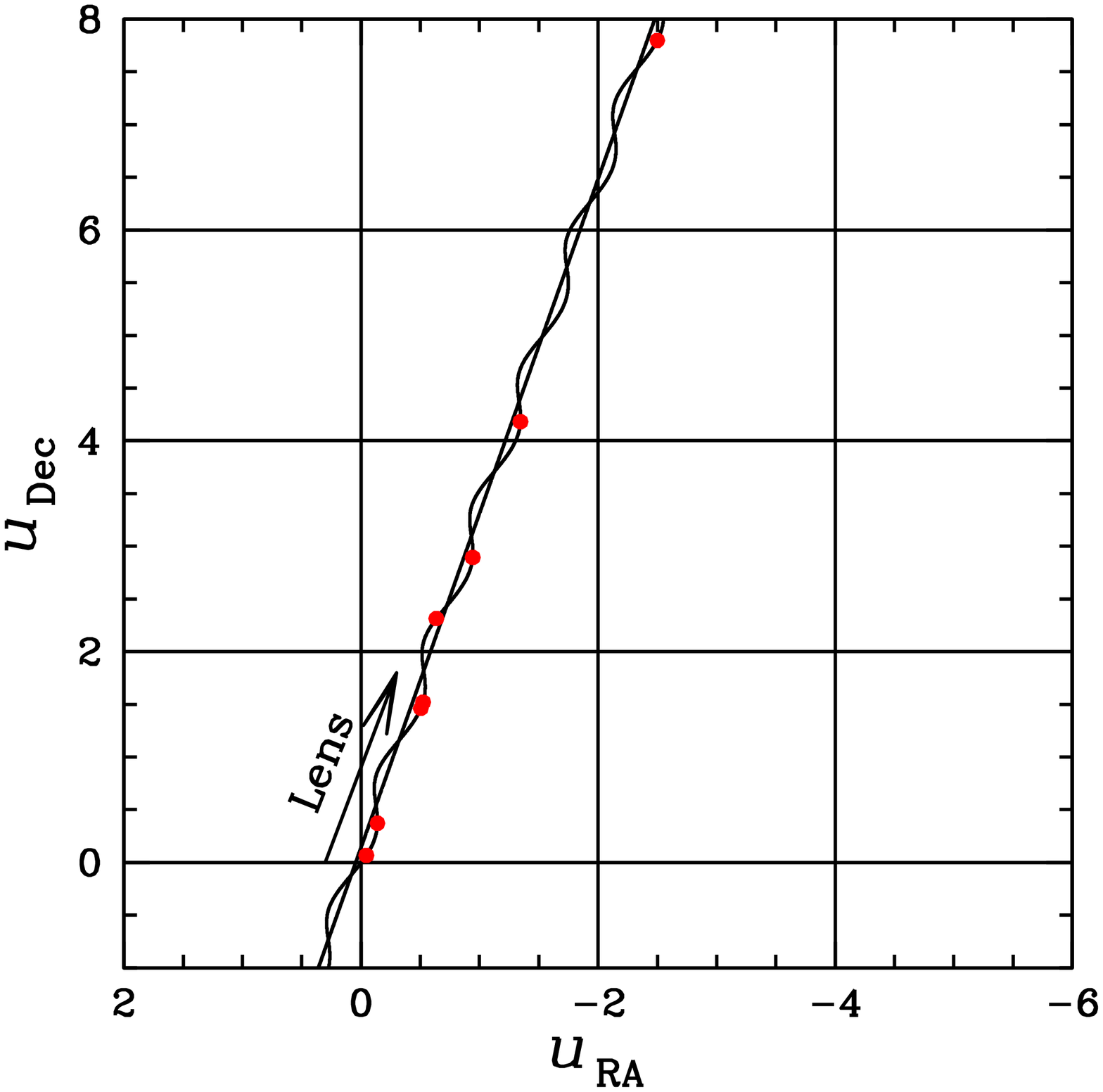}
\plotone{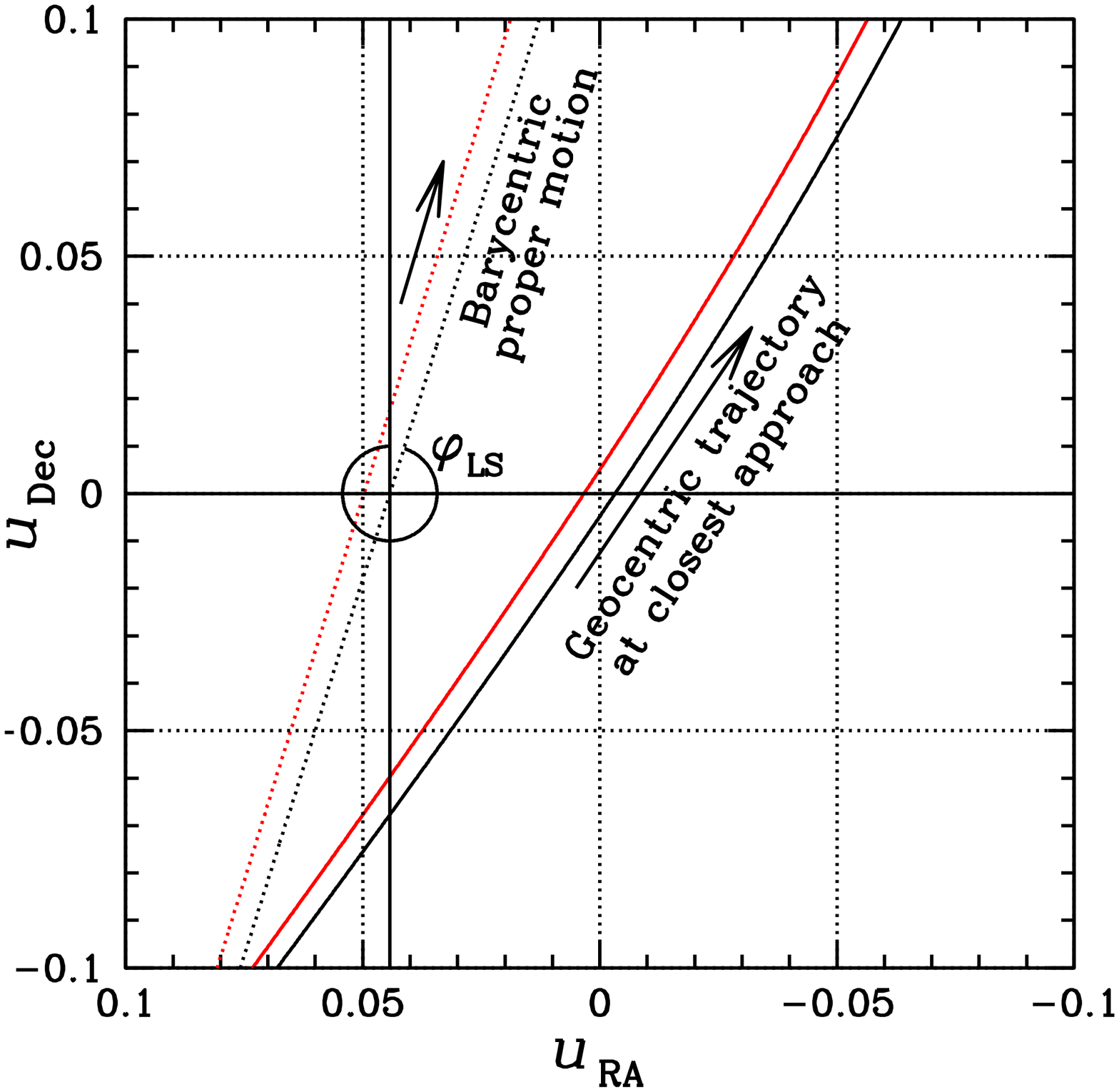}
\caption{Top panel: path of the lens with respect to the source, the position of the source being fixed at (0,0). The direction of motion is shown by an arrow. The black straight line is the lens path without parallax, and the wavy line is the path including parallax. Red points mark the epochs of \HST\/ observations. $u_{\rm RA}$ and $u_{\rm Dec}$ are the RA and Dec components of the lens-source separation, in units of $\theta_\mathrm{E}$. Bottom panel: enlarged view of the lens
trajectory with respect to the source around closest angular approach. Black lines correspond to the $u_{0,-}$ solution, which is the preferred solution, and the red lines show the less-preferred $u_{0,+}$ solution. In both cases, the barycentric trajectories (no parallax), shown by dotted lines, pass on the north side of the source. The geocentric trajectory (i.e., the trajectory as seen 
by an observer on Earth, which includes parallax) of the $u_{0,+}$ solution (shown by the solid red line) passes on the north side of the source, but the trajectory of the $u_{0,-}$ solution (shown by the black solid line) passes to the south. The $u_{0,-}$ solution is used in our analysis presented here. It is worth noting, however, that the derived $\theta_\mathrm{E}$, and hence the mass of the lens, is nearly identical for both solutions.
$\phil$ is the position angle of the barycentric proper motion of the lens with respect to the source (i.e., no parallax), as shown here.}
\label{fig:lens_path}
\end{figure}

In order to understand how astrometry can constrain the direction of motion of the lens, it is useful to consider an illustrative plot of the motion of the lens relative to the source, as shown in Figure~\ref{fig:lens_path}. North is at the top, east on the left; the $(u_{\rm RA},u_{\rm Dec})$ coordinates give the position of the lens relative to the source in units of $\thetaE$, with $u_{\rm RA}$ increasing to the east.  We have used our actual final model described below in
\S{\ref{subsub:second_approach}} for this illustration.
The top panel shows the motion of the lens, with $\piE=0.0894$ and $\phil=342\fdg5$, and an impact parameter of $u_0=0.00271$. The straight line represents the proper motion of the lens with respect to the source, while the wavy line adds the parallactic motion, computed using the JPL ephemeris of the Earth.\footnote{\url{https://ssd.jpl.nasa.gov/horizons/app.html\#/}}
Red dots show the position of the lens at the eight epochs of our \HST\/ observations.

The bottom panel in Figure~\ref{fig:lens_path} shows an enlarged view of the lens trajectory near the source position. The dotted black line represents the proper motion of the lens with respect to the source without parallax, while the solid black line includes the parallactic effect. (The red lines correspond to a
less-preferred $u_{0,+}$ solution described in \S\ref{subsub:lens_motion}.)  The plot shows that near the closest angular approach---and thus the peak \amplification---the relative path is substantially affected by parallax; however, the astrometric deflection is very small at this time (see \S\ref{subsec:astrometric_microlensing}).  Since the source deflection is always in the direction of the line joining the instantaneous position of the lens to the undeflected position of the source, the directions of the source deflections at late times will remain nearly constant with little parallax effect; thus the late-time deflection directions robustly constrain the orientation, $\phil$, of the lens trajectory.

\subsubsection{Constraining the Lens Trajectory Orientation \label{subsubsec:lensorientation} }

Since the parallactic effect is unimportant for constraining $\phil$,
we first fitted the photometry using a light-curve model that neglects parallax. 
The resultant model (with $t_0-2450000 = 5763.33$, $\tE=231.56$~days, $A_{\rm max}=372.62$, and $g=19.3$) predicts the total deflection in units of $\thetaE$ as a function of time, through Equation~(\ref{eq:delta}).

As described in \S\ref{subsec:astrometricanalysis}, we have accurate measurements of the $(x,y)$ positions of the source at the eight epochs of \HST\/ observations. These positions are affected both by the proper motion of the source, and its deflections, which are a function of $\thetaE$ and $\phil$. We fitted a model to the positions, whose parameters are the $x$ (RA) and $y$ (Dec) components of the proper motion, $\thetaE$, and $\phil$.
This fit resulted in values of $\thetaE=5.2 \pm 0.5$ mas and $\phil = 337\fdg9 \pm 5\fdg0 $.

\begin{figure*}
\epsscale{1.10}
\plotone{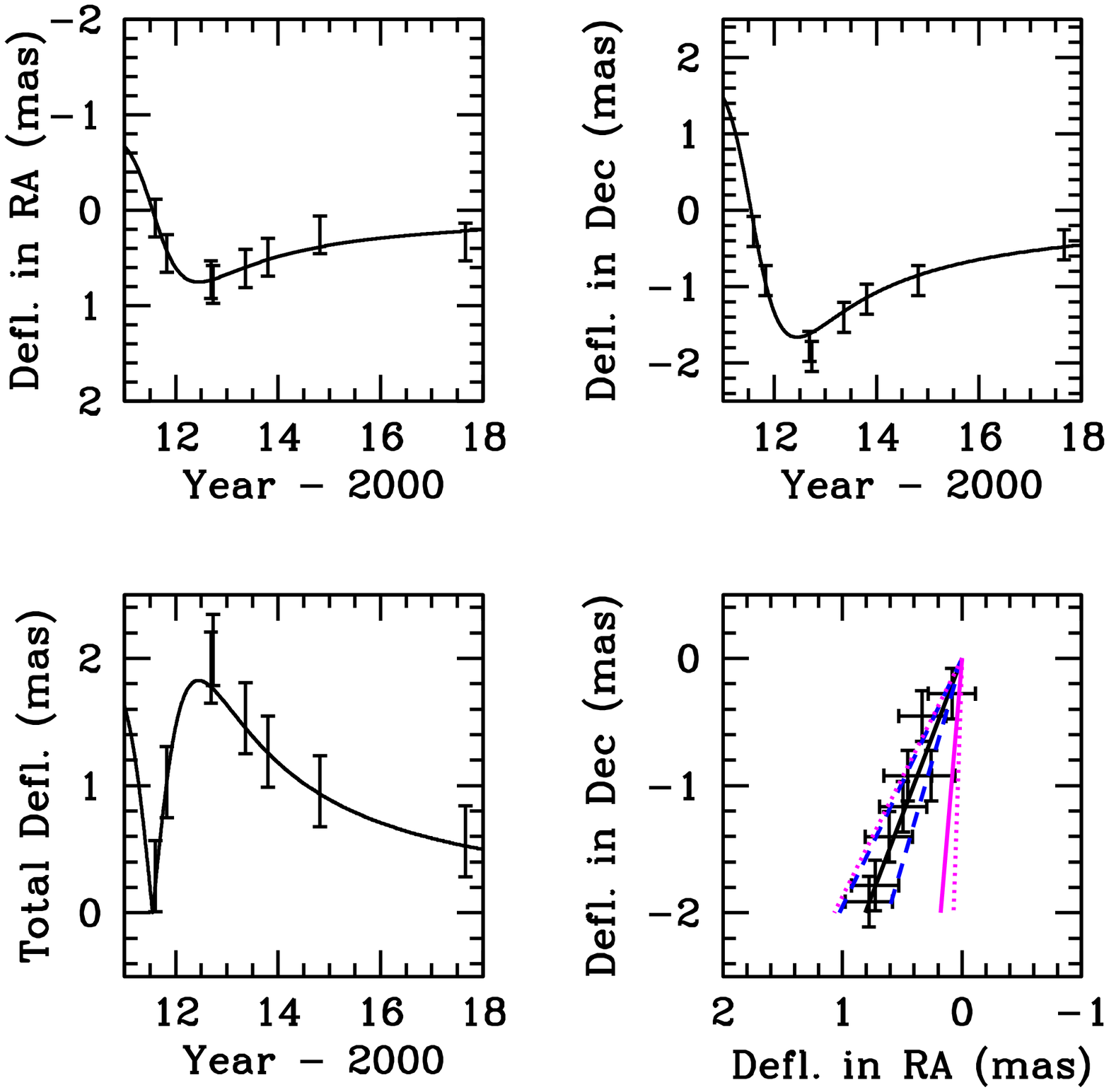}
\caption{ The top two panels show the average values of the measured deflections in RA and Dec at each \HST\/ epoch. We use the values of $t_0, t_\mathrm{E}$, and $u_0$ as derived from a light-curve fit without parallax, and fit for the proper motion of the source, $\theta_\mathrm{E}$, and \phil. The solid black line is the best fit  with $\theta_\mathrm{E} = 5.2$~mas and $\phil=337\fdg9$. The bottom left panel shows the total deflections at each epoch.  The bottom right panel shows the RA versus Dec deflections. These deflections are always along the line joining the lens to the source, and thus at large separations their direction is opposite to the direction of the relative motion of the lens with respect to the source. The solid black line passes through the origin, at the $\phil$ angle derived above. The dotted blue lines indicate the allowed range from astrometry based on the uncertainties. The allowed range of $\phil$ from
photometry alone is shown by the dotted magenta lines, the best fit value being $\phil=354\fdg8$  shown by the solid magenta line. 
The combined constraint from photometry and astrometry is used in our subsequent analysis.}
\label{fig:astrom_simple}
\end{figure*}

Components of the resultant deflections as a function of time, after subtracting the 
best-fitting proper motion, are shown in the top two panels of Figure~\ref{fig:astrom_simple}. These panels plot the deflections in the RA and Dec directions.  The bottom-left panel 
shows the total amount of deflection, again as a function of time. Note that the total deflection reaches a maximum of $\sim$2~mas in late 2012.

The bottom-right panel of Figure~\ref{fig:astrom_simple} plots the RA versus Dec deflections. These deflections are always along the line joining the lens to the source, and thus at large separations their direction is opposite to the direction of the relative motion of the lens. The solid black line passes through the origin, at the $\phil$ angle derived above. The dotted blue lines indicate the allowed range based on the uncertainties. 
Although the best fit value from
photometry alone is $\phil=354\fdg8$ (shown by the solid
magenta line), the allowed range of $\phil$ (shown by the dashed magenta lines)
has a flat probability distribution. 
The combined constraint from photometry and astrometry is used in our subsequent analysis.

Note that for clarity in Figure~\ref{fig:astrom_simple} we have shown the mean deflections at each epoch, and not the individual measurements. In particular, this avoids a confusing overlap of points in the bottom-right panel, where the deflections are not a monotonic function of time. 
The individual measurements are shown in the next section. 


\section{Full Modeling of the Photometric and Astrometric Data}
\label{sec:fullmodeling}

In this section we give full details of our analysis, carried out independently by several coauthors using different parameterizations, all leading to a consistent final model of the event. 

\subsection {First Approach: All Photometric Data Sets, and Robustness of Parallax Measurement}
\label{subsub:other_photometry}

In addition to OGLE and MOA, photometric time-series data were obtained at several more observatories (see Table~\ref{table:journal}).  These data typically cover a relatively narrow range of $\sim$20 days around the peak, beíng primarily aimed at searching for planet-related distortions of the light curve. These photometric series, unlike those obtained by the survey programs, do not provide significant constraints on the lens-source model---especially since each set of observations can have a different baseline magnitude and blending parameter.  Nevertheless, in our initial approach, we included all the data in our analysis, but also carried out analyses separately for ``OGLE-only,'' ``MOA-only," and ``OGLE+MOA-only'' data sets.

Ground-based time-series photometry is susceptible to systematic noise, and this must not be mistaken for real features of the light curve.
To improve the robustness of our solution, 
we model the photometric uncertainties, and moreover force the model to follow the bulk of the data by 
explicitly down-weighting outliers. 
As implemented for the SIGNALMEN microlensing anomaly detector \citep{Dominik2007,Dominik2019}, we specifically adopt a bi-square weight function with regard to the median residual, and a Gaussian distribution for the uncertainties with revised standard deviation in magnitude of
\begin{equation}
\tilde{\sigma} = \sqrt{(\kappa\,\sigma)^2+\sigma_0^2} \, ,
\end{equation}
where $\sigma$ denotes the reported error bar, $\kappa$ is a scaling factor, and $\sigma_0$ corresponds to a systematic error added in quadrature.
For the plot of the various data sets as shown in Figure~\ref{fig:AllPhotDataPlot}, we give the respective estimated values of $\kappa$ and $\sigma_0$ in Table~\ref{Tab:ErrorBars}.
If the size of the error bars does not vary substantially, there is a degeneracy between $\kappa$ and $\sigma_0$, and either of the parameters
provides modified constant error bars (while it does not matter which). 

We find two viable models, significantly only distinguished by the sign of $u_0$, and will refer in the following to the model with $u_{0,-}$.
The microlensing parallax parameter $\pi_\mathrm{E}$ is constrained by the wing of the light curve and much less sensitive to the peak region,
and therefore is mostly constrained by the microlensing survey data. 
From various combinations of data sets, we consistently find $\pi_\mathrm{E} = 0.10 \pm 0.02$, but some variation in the trajectory angle $\psi$, correlated with $t_\mathrm{E}$ and the blend fraction, yielding visually indistinguishable model light curves.

However, the angle of lens-source proper motion, $\phil$, follows robustly from the astrometric data (see \S\ref{subsubsec:astrometricconstraints}), given that the centroid shift to first order (i.e., neglecting the small distortion caused by parallax) traces an ellipse (a highly flattened ellipse resembling a line in our case) whose semi-major axis is parallel to $\vec{\mu}_\mathrm{LS}$. If we restrict this angle to the range $333^\circ \leq \phil \leq 343^\circ$, as suggested by the astrometric data, the photometric light curve does not change substantially, and we find $\pi_\mathrm{E} \simeq 0.086$.

We emphasize here that it is incorrect to say that there is a discrepancy between
the paths determined from photometry and astrometry, since 
there is a correlation between $\phil$ and $\pi_\mathrm{E}$
in the photometric solution.
However, restricting the trajectory angle $\phil$ as robustly derived in the last section
from the orientation of the centroid shifts to the range $333^\circ \leq \phil \leq 343^\circ$, suggests $\pi_\mathrm{E} \simeq 0.086$. 

 \subsection {Second Approach: Simultaneous Fit of Photometric and Astrometric Data} 
 \label{subsub:second_approach}
We now turn to a full analysis in which we fit the astrometric and photometric data simultaneously in order to obtain all of the parameters. Such a solution is important, since the crucial parameters of $\thetaE$ and $\piE$ are derived from two different types of data. 
A simultaneous solution is also essential for a correct estimate of the uncertainties in the model parameters. 

We follow the same plane-of-the-sky approach described in \S\ref{subsec:heuristic}, which makes it easier to work with, and also show the actual paths of the lens and the source, and the deflections.
We follow  a different parameterization procedure where the model parameters we optimize contain all terms needed to characterize the positions of the lens and the source on the sky as a function of time; these include the reference positions and proper motions of both lens and source, their relative parallax, and the angular Einstein radius of the lens.  In principle, the source parallax is also needed; however, its parallax in the reference system we use is close to zero, and it is not meaningfully constrained by the observations.  As discussed below in \S\ref{subsec:sourcedistance}, the best constraints on the source distance come instead from photometry and high-resolution spectroscopy.

From these parameters, the undeflected paths of the source and lens can be determined. The deflection of the lensed image of the source is then computed, and the resulting deflected source positions are matched to the observed positions.  The same calculation also yields the source magnification; in order to match the observed photometry, the model must include a baseline magnitude of the source and a blending parameter for each photometric data set.  Consistent with the previous approach, we found that most of the photometric data sets cover too short a time interval to yield meaningful constraints on the event parameters in the presence of significant blending; therefore we limit the model optimization to the MOA and OGLE photometric data sets, and validate the resulting model for the other data sets separately (see \S\ref{subsub:other_photometry}).   Also, in order to avoid undue impact from any secular variations in photometric responses, we only include OGLE and MOA photometric measurements within $ \pm 2 $ years from the peak of the event.  We adopt the approximate values of \phil, $t_\mathrm{E}$, $t_0$, and $u_0$ from the analysis of the previous section as our initial estimates, but we leave all parameters free in the optimization; the baseline magnitude and blending parameters for MOA and OGLE are also separately optimized.

As discussed in the previous subsection, the results of the optimization depend to some extent on the relative weighting of astrometry and photometry.  Because the number of photometric measurements greatly exceeds that of astrometric measurements, and the nominal photometric uncertainties are very small, an optimization using nominal errors disproportionately weights photometry, resulting in a poor match to the astrometry.  
In order to obtain a more balanced weighting of astrometry and photometry,
we scaled the photometric errors by different amounts in 
different temporal bins making sure that the scaled errors are compatible with the statistical dispersion in the measurements, and validated each solution based on a reasonable match to the astrometric data. In our final model, most photometric points are still at an uncertainty below 10~mmag.  This solution has a total astrometric $ \chi^2 $ of 136 with 106 points. (The solution with nominal weights has a higher astrometric $ \chi^2 $ of 149.)  The parameters of the final model are given in Table~\ref{table:parameters_SC}.

\begin{deluxetable*}{lcDDc}
  \tablecaption {Parameters of the Full Fit to Astrometry and
    Photometry\label{table:parameters_SC}}
  \tablehead{
    \colhead{Parameter} &
    \colhead{Units} &
    \multicolumn2c{Value} &
    \multicolumn2c{Uncertainty (1$\sigma$)} &
    \colhead{Notes\tablenotemark{a}}
  }
\decimals
\startdata
$ \mu_{\mathrm S} $ (RA)      & $\masyr$  &    -2.263  &   0.029   & (1)  \\
$ \mu_{\mathrm S} $ (Dec)     & $\masyr$  &    -3.597  &   0.030   & (1)  \\
$ \theta_{\mathrm E} $        & mas      &     5.18      &   0.51    & (2) \\
$ t_{\mathrm E}^{\mathbf{\star}} $ & days &  270.7       &  11.2     &  (3)\\
$ \phil$                      & deg      &   342.5       &   4.9     & (4)  \\
$ t_0^\star $ (HJD$-$2450000.0) & days   &  5765.00      &   0.87    & (5)  \\
$ \pi_{\mathrm E} $           &          &     0.0894    &   0.0135    & (6)  \\
$ u_0^\star $        &          &     0.0422    &   0.0072    & (7)  \\   
MOA  Baseline $ R $ magnitude & mag      &    16.5147    &   0.0016   & \\
MOA Blending parameter        &          &    16.07      &   0.66    & \\
OGLE Baseline $ I $ magnitude & mag      &    16.4063    &   0.0015  &  \\
OGLE Blending parameter       &          &    18.80      &   0.79    &  \\
\noalign{\vspace{12pt}}
~~~Derived parameters:\\
$ {u_0} $              &          & 0.00271 & . & (8)\\
$ {t_0} $  {(HJD$-$2450000.0)} & {days} &
                            5763.32 & . & {(9)}\\
\enddata
  \tablenotetext{a}{Notes: (1) Undeflected proper motion of the source in \Gaia\/ EDR3 absolute frame; 
  (2) Angular Einstein radius;
  (3) Angular Einstein radius $\thetaE$ divided by absolute value of lens-source proper motion $\mu_\mathrm{LS}$;
  (4) Orientation angle of the lens proper motion relative to the source (N through E);
  (5) Time of closest angular approach without parallax motion;
  (6) Relative parallax of lens and source in units of $ \theta_\mathrm{E} $; 
  (7) Impact parameter in units of $ \theta_\mathrm{E} $ without parallax motion;
  (8) Impact parameter derived using the model-fit parameters above and after including parallax motion, in units of $\theta_\mathrm{E}$;
  (9) Time of closest angular approach derived as above and after including parallax motion.
  }
\end{deluxetable*}

\begin{figure*}
\begin{center}
    \includegraphics[width=5in,angle=90]{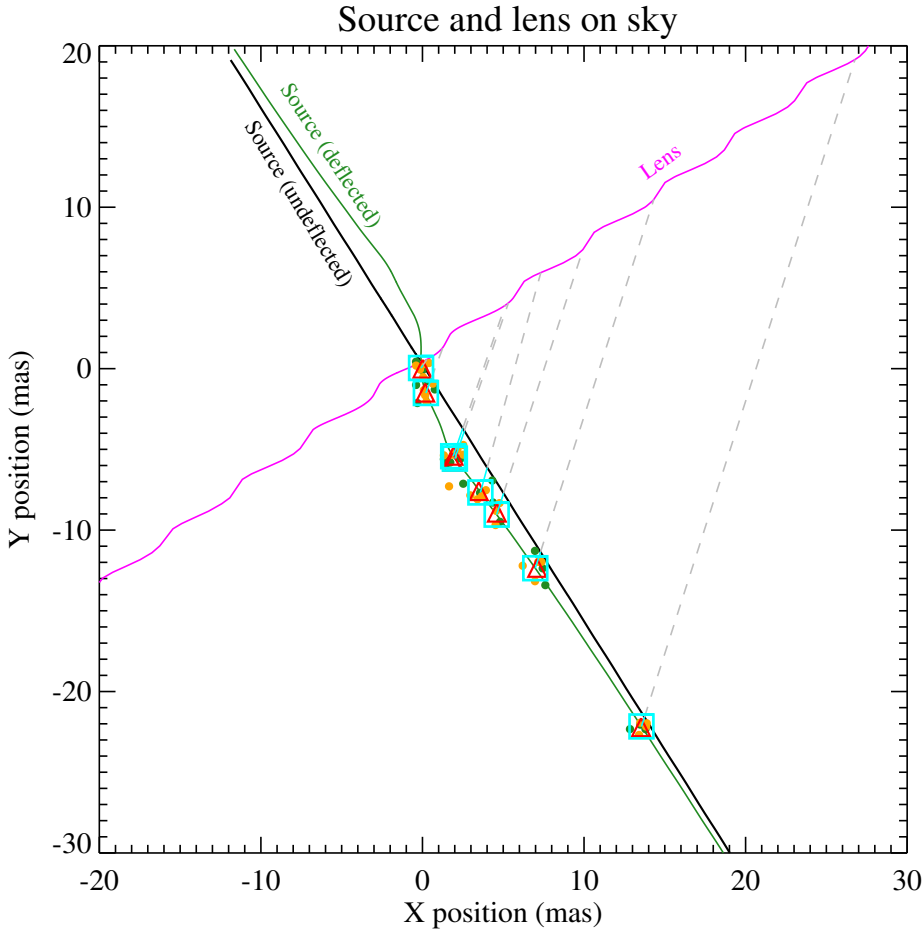}
\end{center}
    \caption{Representation of the reconstructed motions of the lens and of the source on the plane of the sky.  The reference point is the undeflected position of the source at time $ t_0 $.  The small dots are the individual measurements from {\HST\/} images (green for F606W, orange for F814W); the larger red triangles represent the average positions for each epoch.  The cyan squares are the fitted positions at each epoch.  The gray lines connect the undeflected source and lens positions at each {\HST\/} epoch.}
    \label{fig:plane_of_sky}
\end{figure*}

Figure~\ref{fig:plane_of_sky} shows the reconstructed motion of the lens (magenta) and of the source (black) in the plane of the sky, based on our adopted model.  The predicted apparent source trajectory is shown by the green solid line. The astrometric measurements are shown individually by the small filled circles, and their epoch averages as red triangles.  Cyan squares show the model position at each epoch; gray lines connect the model lens position to the undeflected and deflected source positions at each epoch.  

\begin{figure*}
\begin{center}
    \includegraphics[width=5.6in,angle=90]{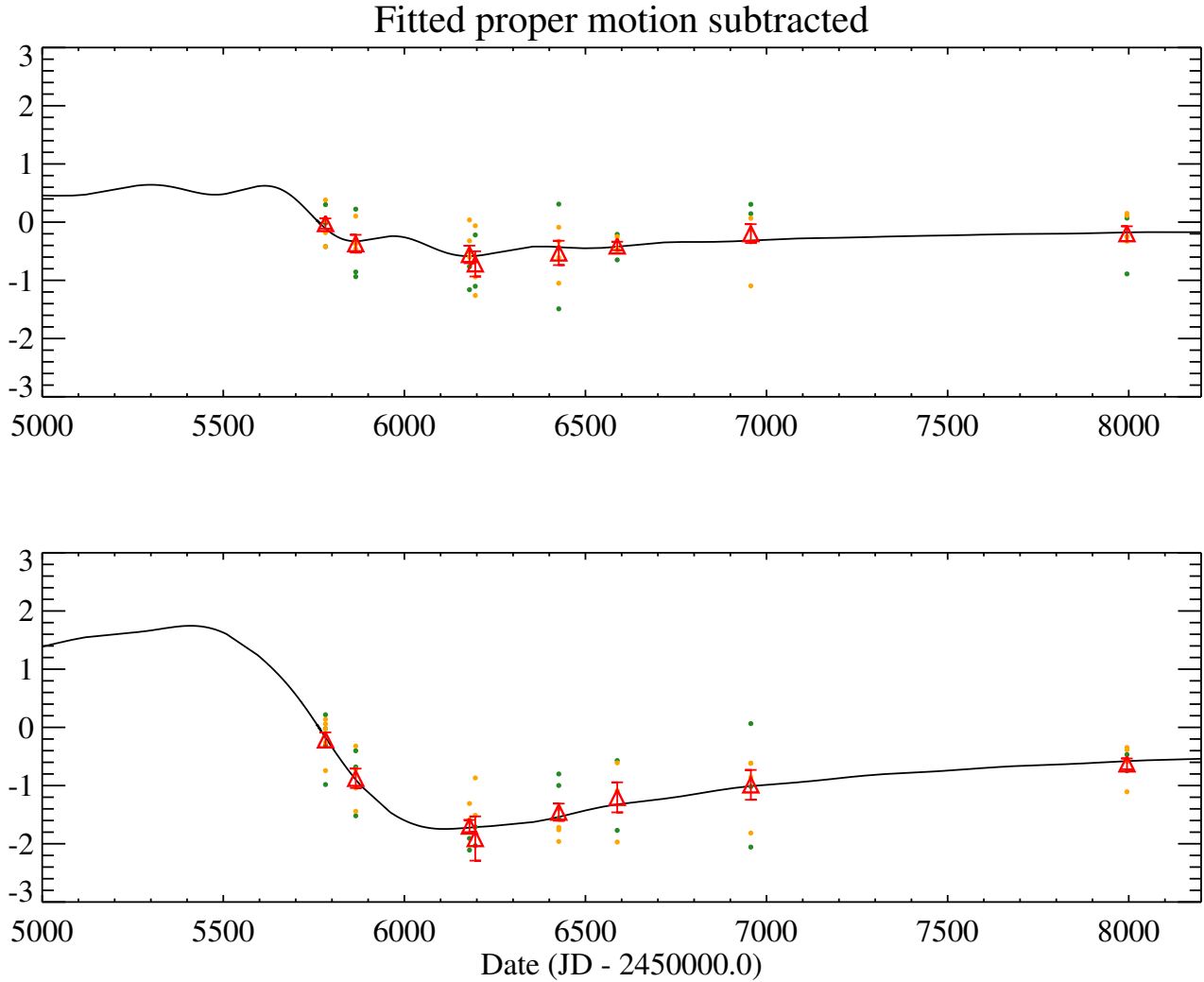}
\end{center}
    \caption{Predicted and measured positions of the source for our adopted joint astrometric and photometric fit.  Small dots show the individual measurements from {\HST\/} images (green for F606W, orange for F814W), while red triangles with error bars show the average and uncertainty at each \HST\/ epoch.  The black line is our final adopted model.
    The fitted proper motion for the source has been subtracted from both model and measurements, in order to allow a better scaling of the plot.}
    \label{fig:xy_deflections_vs_time}
\end{figure*}

Figure~\ref{fig:xy_deflections_vs_time} presents the measured and predicted source positions separately for the $ x $ and $ y $ coordinates. To improve the legibility of the plot, the fitted proper motion of the source has been subtracted from both model and data; therefore the points shown represent the deflection of the source.  
The black line is our final adopted model which takes 
photometry as well as astrometry into account.

We can now check the consistency of the final model with our measured \HST\/ photometry
in the F606W and F814W filters, and also constrain the amount of blending.
We used the final model to calculate the \amplification s at the \HST\/ observation epochs, and varied the baseline magnitudes and the blending factor 
to fit the \HST\/ photometry.
The resulting fit is excellent (see Figure~\ref{fig:HST_lightcurve}), and
yields baseline source magnitudes of 
$m_{\rm F606W} = 21.946 \pm 0.012$ and $m_{\rm F814W} = 19.581 \pm0.012$, with 
corresponding blending factors for the \HST\/ photometry of 
$g= -0.012 \pm 0.015$ and $-0.006 \pm 0.012$, respectively.
The bottom two panels of Figure~\ref{fig:HST_lightcurve}
show the residuals of the observations relative to the model in F606W and F814W at the epochs of \HST\/ observations,
where we have assumed the minimum physically allowed value of $g=0$. 
(It makes little difference if we instead use the slightly negative values
of $g$ from the formal fit.) The stringent constraints on blending and
lack of color variation make it unlikely that our deflection measurements could be
affected by blending with a binary companion or a
field star lying within the \HST\/ PSF.

\begin{deluxetable*}{lcc}[bt]
\tablecaption{Revised Error Bars of Photometric Data \label{Tab:ErrorBars}}
\tablehead{
\colhead{Data set} &
\colhead{$\kappa$} &
\colhead{$\sigma_0$}
}
\startdata
OGLE          & 1.25 &  0.005 \\
MOA           &  1.18  & 0.0025 \\
Wise Survey & 0.64& 0.009\\
Danish 1.54m DFOSC    & 2.75 & $10^{-5}$ ($^\ast$)\\
Danish 1.54m LuckyCam      &0.1  ($^\ast$)  & 0.050 \\
MONET North 1.2m      & 0.57 & 0.006 \\
Faulkes North 2.0m         &    0.1  ($^\ast$)       & 0.014 \\
Liverpool 2.0m          &    1.17          & 0.008 \\
SAAO 1.0m $I$    &  1.42 &  0.0014 \\
SAAO 1.0m $V$   & 0.59 & 0.009 \\
Tasmania 1.0m       & 0.28& 0.017 \\
CTIO 1.3m $I$ & 0.62 & $10^{-5}$ ($^\ast$) \\
CTIO 1.3m $V$ & 1.01  & $10^{-5}$ ($^\ast$) \\
Auckland 0.4m &  0.91  & 0.007\\
Farm Cove 0.35m & 1.00 & 0.003 \\
Kumeu 0.35m & 1.87 &$10^{-5}$ ($^\ast$)\\
Vintage Lane 0.4m & 0.1  ($^\ast$) & 0.011 \\
{Weizmann 0.4m} & 0.80 & $10^{-5}$ ($^\ast$) \\
{Wise 0.46m} & 0.22 & 0.009
\enddata
\tablecomments{The adopted error bar (in magnitude) becomes $\tilde{\sigma} = \sqrt{(\kappa\,\sigma)^2 + \sigma_0^2}$, where $\sigma$ is the reported error bar of the photometric data. We have applied range constraints $\kappa > 0.1$ and $\sigma_0 \geq 10^{-5}$. An asterisk ($^\ast$) 
indicates that the value is at the range boundary. }

\end{deluxetable*}

\subsection{Lens Motion}
\label{subsub:lens_motion}
The path of the lens in the sky plane, as derived from the above analysis, is shown in
Figure~\ref{fig:lens_path}.  We note that there are two solutions corresponding to 
$u_{0,+}$ and $u_{0,-}$ \citep{Gould2004}.
However, at $t_0$, the angular separation between the source and the lens is dominated by the parallactic motion of the lens (the separation caused by the parallactic motion is $\sim$0.04, compared to the actual value of $u_0 \simeq 0.00271$). Thus, the path without parallax in both solutions lies on the $u_{0,+}$ side. 

The paths of the lens for the $u_{0,+}$ and $u_{0,-}$ solutions are shown in Figure~\ref{fig:lens_path}, where the dashed lines show the path of the lens without parallax, and the solid lines show the path of the lens after the parallactic motion is taken into account. The respective paths are separated only by 0.02 mas at the time of maximum {\amplification}, and quickly merge.  The $u_{0,-}$ solution is the preferred solution, and is the one used in our analysis. However, we verified that both solutions provide practically identical deflections (since the deflection measurements are at much higher $u$, where the two paths nearly merge), and the results  are the same for all practical purposes in both solutions.

The first two rows of Table 4 give the proper motions of the source. We use the values of 
$\thetaE$, $t_{\rm E}^\star$, and \phil\ given in the next three rows to determine the
proper motion of the lens with respect to the source as $-2.10$ $\pm$ 0.22 and 6.66 $\pm$ 0.67 mas yr$^{-1}$ in RA and Dec, respectively. The resulting absolute proper motions of the lens 
are given below in Table~\ref{table:lensproperties}.

\section{Properties of the Source \label{sec:source_properties} }

As described in \S\S\ref{subsec:photometric_microlensing}--\ref{subsec:astrometric_microlensing}, the mass determination for the lens does not depend upon the individual distances to the lens and source, but only on the relative lens-source parallax, $\piLS$, and the Einstein ring radius, $\thetaE$---quantities that are directly determined from the light curve and the measured astrometric deflections.  However, we still need an estimate of the distance to the source, $\DS$, in order to determine the distance to the lens, $\DL$, which is discussed below.  Moreover, \OB\ was a very high-magnification event, with an impact parameter of only $u_0\simeq0.00271$. Thus it is desirable to estimate the angular diameter of the source, to verify that it is consistent with the point-source light-curve modeling adopted in the previous section.

In this section we use results of ground-based spectroscopy of the \amplified{} source, and \HST\/ photometry at its baseline brightness, to estimate its distance and angular diameter.

\subsection{High-Resolution Spectroscopy}

In a target-of-opportunity program focusing on high-dispersion spectroscopy of high-amplitude microlensing events in the Galactic bulge, \citet{Bensby2013} obtained four spectra of \OB\ on 2011 July 20 and 21. These observations were made almost precisely around the dates of maximum magnification. The authors employed spectrographs on three different large telescopes: VLT, Magellan, and Keck~I, at spectral resolutions ranging from 46,000 to 90,000. The aim of their program was to determine chemical compositions of dwarfs and subgiant stars in the Galactic bulge.

\citet{Bensby2013} carried out a non-LTE model-atmosphere analysis of the spectra, obtaining an effective temperature and surface gravity of $\Teff=5382\pm92$~K and $\log g=3.80\pm0.13$, consistent with a late G-type subgiant. The metallicity was found to be slightly above solar, at $\rm[Fe/H]=+0.26\pm0.14$, and the radial velocity is $+134.0\,\kms$. 

As discussed above (e.g., Figures~\ref{fig:8x8zoomin} and \ref{fig:postagestamps}), the \HST\/ images show that \OB\ is accompanied by a neighboring star only $0\farcs4$ away, which would have been included in the spectrograph apertures.   However, at the dates of the observations, \OB\ was $\sim$3.2~mag brighter than the neighbor in the F606W bandpass; thus the contamination was about 5\%, which should have a minimal effect on the spectroscopic investigation.

As shown in Figure~\ref{fig:CMD}, the neighbor has a color similar to that of the baseline source, but is more luminous.
We verified its small effect on the spectroscopy by artificially contaminating the source spectrum with a 5\% flux from a giant star with the same effective temperature and metallicity, but with $\log g=0$. This produced a smaller change in the line strengths of the combined light than would a change of $\log g$ for the source star by its 0.13~dex uncertainty.

\smallskip
\subsection {Distance and Angular Diameter \label{subsec:sourcedistance} }

\begin{figure}
\begin{center}
\includegraphics[width=0.46\textwidth]{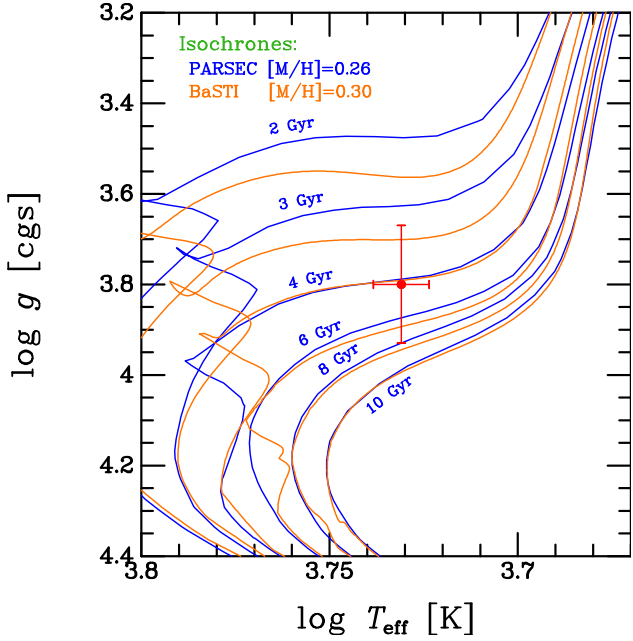}
\end{center}
\caption{
Location of the \OB\ source star in the $\log g$ versus effective-temperature diagram (red filled circle with error bars), using stellar parameters from \citet{Bensby2013}. Superposed are theoretical isochrones from two sources. The blue curves show PARSEC isochrones, calculated for the source's measured metallicity of $\rm[M/H]=+0.26$, for ages of 2 to 10~Gyr. The orange curves are BaSTI isochrones for $\rm[M/H]=+0.3$ and the same range of ages.
\label{fig:padova}
}
\end{figure}

We estimate the distance to the source using two independent sets of theoretical stellar isochrones: ``Padova''  isochrones, which we obtained using the PARSEC web tool\footnote{Version 3.6 at \url{http://stev.oapd.inaf.it/cmd}} \citep{Bressan2012,Marigo2017}; and ``BaSTI'' isochrones\footnote{\url{http://basti-iac.oa-abruzzo.inaf.it/index.html}} \citep{Hidalgo2018,Pietrinferni2021}.  In Figure~\ref{fig:padova} we plot the position of the source in the distance-independent $\log g$ versus $\log\Teff$ plane (red filled circle with error bars). Superposed are PARSEC isochrones, calculated for the measured metallicity of $\rm[M/H]=+0.26$ (blue curves), and BaSTI isochrones, for a metallicity of $\rm[M/H]=+0.3$ (orange curves). Both sets of isochrones agree well around the location of the source star, yielding an age of about 4~Gyr, but with considerable uncertainty.

At the position of the source in the diagram, a star on the PARSEC 4-Gyr isochrone has absolute magnitudes (Vegamag scale) in the native \HST/WFC3 bandpasses of $M_{\rm F606W}=+3.04$ and $M_{\rm F814W}=+2.40$, and an intrinsic color of $(m_{\rm F606W}-m_{\rm F814W})_0 = 0.64$. The BaSTI models---which have been updated recently, using new filter throughput curves released in 2020 October by the WFC3 team \citep{Calamida2021}---yield similar absolute magnitudes of +3.01 and +2.38, and an intrinsic color of 0.63. Taking the averages, and using the observed baseline color of $m_{\rm F606W}-m_{\rm F814W} = 2.365$ (Table~\ref{table:basicdata}), we find a color excess of $E(m_{\rm F606W}-m_{\rm F814W})=1.73$. Based on the values of $A_\lambda/A_V$ given for the WFC3 filters at the PARSEC website (derived from the interstellar extinction curves of \citealt{Cardelli1989} and \citealt{O'Donnell1994}, with $R_V=3.1$), this color excess corresponds to an extinction of $A_{\rm F814W}=3.33$~mag. Thus, using an apparent magnitude at baseline of $m_{\rm F814W}=19.581$ from Table~\ref{table:basicdata}, we find a true distance modulus of $(m-M)_0=13.86$, giving a linear distance of $D_S=5.9\pm1.3$~kpc. 
The quoted error is the formal uncertainty (dominated by the uncertainty in the spectroscopic $\log g$), but there likely are additional systematic errors, given the large amount of extinction, the assumption of a standard value of $R_V$, the high sensitivity to $\log g$, and other sources. For example, if a low value of $R_V\simeq2.5$, as deduced by \citet{Nataf2013} for the Galactic bulge region, were used, the distance would increase by $\sim$2.8~kpc.

The nominal 5.9~kpc distance places the source on the near edge of the Galactic bulge. However, it cannot be ruled out that the source is actually located closer to the center of the Galactic bulge, at a distance of $\approx$8~kpc. The star's high radial velocity is consistent with stars in the bulge, but not large enough to rule out disk membership.

Taking the spectroscopy, photometry, and derived distance at face value, these parameters correspond to an angular diameter of the source of $\sim$0.0038~mas.

\subsection{{Finite-source Effects Are Negligible}}
\label{sec:finite}

We can estimate the effect of the finite size of a small source on the light curve by
expanding $A(u)$ around $u = u_0$, giving
 \begin{eqnarray}
   A(u) & \simeq & A(u_0) + A'(u_0)\,(u-u_0) + \nonumber \\
   & & \quad +\, \frac{1}{2}\,A''(u_0)\,(u-u_0)^2\,,
 \end{eqnarray}
where the prime symbol denotes the derivative. If we average over the face of a spherically symmetric source star, the term with the first derivative cancels out, leaving the quadratic term as the first one 
that contributes to the difference between finite and point-like source.
Averaging the latter over an extent $\delta$ explicitly gives
\begin{eqnarray}
(\Delta A)_\delta(\delta) & = & \frac{1}{4\delta}\,A''(u_0) \int\limits_{u_0–\delta}^{u_0+\delta} (u-u_0)^2\,\mathrm{d}u \nonumber \\
& = & \frac{1}{6}\, A''(u_0)\,\delta^2\,.
\end{eqnarray}

For $u \ll 1$, one finds $A(u) \simeq 1/u$ and thereby $A''(u) \simeq 2/u^3$. Consequently, 
we find the largest differences between finite and point-like source for the smallest $u$.
For a source of angular radius $\theta_\star$, we find a source size parameter
$\rho_\star \equiv \theta_\star/\theta_\mathrm{E}$, and if $u_0 \gg \rho_\star$,
we can approximate the magnification difference by
\begin{eqnarray}
\Delta A & \simeq &\frac{1}{\rho_\star} \int\limits_0^{\rho_\star} (\Delta A)_\delta\left(\sqrt{\rho_\star^2-\eta^2}\right)\,\mathrm{d}\eta \nonumber\\
& = & \frac{1}{6}\, A''(u_0)\,\rho_\star^2 \int\limits_0^1 (1-\xi^2)\,\mathrm{d}\xi \nonumber \\
& \simeq & \frac{2}{9}\,\frac{\rho_\star^2}{u_0^3}\,. 
\label{eq:mdiff}
\end{eqnarray}

With the angular diameter of the source $2\,\theta_\star \simeq 0.0038~\mbox{mas}$ and the angular Einstein radius $\theta_\mathrm{E} = 5.2~\mbox{mas}$, we find $\rho_\star \equiv \theta_\star/\theta_\mathrm{E} = 0.00037$ while $u_0 = 0.00271$. 
Inserting into Equation~(\ref{eq:mdiff}) gives $\Delta A = 1.5$ as compared to $A(u_0) = 369$ at closest angular approach between lens and source, in close agreement
with a numerical evaluation of the average of the exact magnification function $A(u)$.
With a blend ratio $g \simeq 20$, comparing $2.5 \log[(A+g)/(1+g)]$ for $A$ and $A+\Delta A$ gives a difference of 4~mmag between finite source and point-like source, which is of the order of the systematic errors of the photometry. 

Accounting for the finite source size will 
thus result in a very small change in $u_0$, which has
very little effect on other parameters, 
evidenced by their practically identical values for our $u_{0,+}$ and $u_{0,-}$ solutions. Similar to the case of the long-duration microlensing event OGLE-2014-BLG-1186 discussed by \citet{Dominik2019}, the measurement of the parallax parameter $\pi_\mathrm{E}$ comes from the wings of the photometric light curve, not from the peak. 
Additionally, none of the astrometric deflection measurements are close to the peak.
With the finite size of the source making little difference, we can safely neglect any limb-darkening effects.

\section{Nature of the Lens}

In this section we discuss the nature of the lensing object of the \OB\ microlensing event. We consider the mass of the lens, discuss constraints on its optical luminosity, and consider whether it is a single object or could be a binary system. 

\subsection{Mass \label{subsec:mass} } 

As shown by Equation~(\ref{eq:mass}), the mass of the \OB\ lens can be determined from the values of its angular Einstein radius, $\theta_{\rm E}$, and the relative lens-source parallax, $\pi_{\rm E}$. In \S\ref{sec:fullmodeling}, we derived $\thetaE = 5.18\pm0.51$~mas and $\piE = 0.089\pm0.014$. These yield a mass of $M_{\rm lens}= 7.1 \pm 1.3\, M_\odot$.

An object with a mass this large cannot be a single (or double) NS or white dwarf. It can only be a BH---or an ordinary star, or conceivably a binary (or higher multiple) containing stars, BHs, and/or other compact companions. To distinguish between these possibilities, we consider the observational constraints on the luminosity and binarity of the lens. 

\smallskip

\subsection{Lens Distance \label{subsec:lensdistance} } 

The mass determination for the lens is independent of assumptions about the distances to the lens and source. However, to constrain the {\it optical luminosity\/} of the lens, we first do need to determine
its distance. This can be obtained from Equation~(\ref{eq:distances}), using the value of $\piLS = 0.463 \pm 0.051$~ mas (from \S\ref{sec:fullmodeling}), and the distance to the source (from \S\ref{subsec:sourcedistance}), $\DS=5.9\pm1.3$~kpc. 

These values yield a lens distance of $\DL= 1.58\pm0.18$~kpc. Note that the derived distance of the lens is only a weak function of the adopted source distance. If the source were at the distance of the Galactic center, the lens distance would only increase to $\sim$1.7~kpc.

\smallskip

\subsection{Luminosity Constraints}

\subsubsection{From the Baseline Magnitude}

Main-sequence stars with a mass in the range $7.1 \pm 1.3\, M_\odot$  have a range of $I$-band absolute magnitudes of $M_I=-1.5\pm0.3$, based on the tabulation\footnote{Online version of 2021 March 2, at \url{http://www.pas.rochester.edu/~emamajek/EEM_dwarf_UBVIJHK_colors_Teff.txt}} of stellar properties
compiled by \citet{Pecaut2013}. 

The lens distance of $1.58\pm0.18$~kpc, from the previous subsection, corresponds to a true distance modulus of $(m-M)_0 = 11.0\pm0.2$. Thus a main-sequence lens of the measured mass would have an unreddened apparent magnitude of $I \simeq 9.5$. The $I$-band extinction of the source (from \S\ref{subsec:sourcedistance}) is $\sim$3.3~mag, which is likely an overestimate for the lens, since it is considerably nearer than the source and may suffer less extinction. Nevertheless, this amount of extinction gives an expected apparent magnitude of a main-sequence lens of about $I\simeq12.8$, or brighter if the extinction is lower. The star would be yet brighter if it has evolved off the main sequence to a higher luminosity. Since the baseline F814W magnitude of the source (plus lens) is $\sim$19.6 (Table~\ref{table:basicdata}), a main-sequence or brighter stellar lens is conclusively ruled out.

\subsubsection{Direct Limits from Final-epoch \HST\/ Imaging}

A much stronger constraint on its optical luminosity comes from the fact that the lens was not detected in our final-epoch \HST\/ frames. The parameters $\thetaE = 5.18\pm0.51$~mas and $\tE=270.7\pm 11.2$~days imply that the proper motion of the lens with respect to the source is $6.99\pm0.75\rm\,mas\,yr^{-1}$.  At peak magnification in 2011 July, the separation between lens and source was $\sim$0.01~mas. Therefore, at the epoch of our last \HST\/ observation in 2017 (6.1~years after the peak), the lens-source separation had increased to $\sim$42.6~mas. At this large a separation, a luminous lens would cause the PSF of the source in our images to show clear signs of elongation. In order to search for such a distortion, we subtracted a ``library'' F814W PSF\footnote{From \url{https://www.stsci.edu/hst/instrumentation/wfc3/data-analysis/psf}} from the final-epoch F814W images (in the individual un-resampled {\tt \_flc} frames).  We saw no indication of elongation.

\begin{figure*}
\begin{center}
\epsscale{1.17}
\plotone{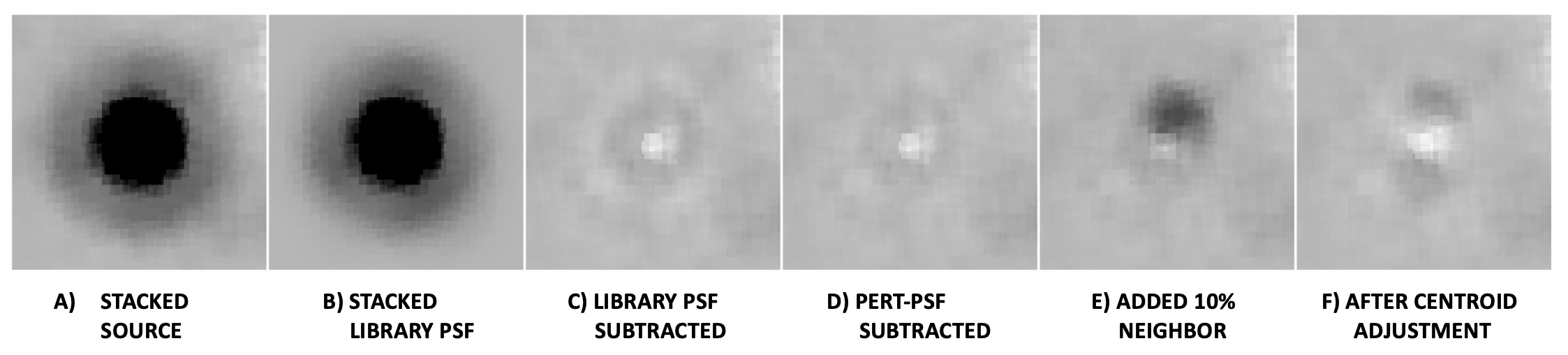}
\end{center}
\caption{
Panels showing 10$\times$10-pixel stacks in the reference frame of 4$\times$250-s F814W images from the 2017 epoch, centered on the source.  (A) The original {\tt \_flc} data with only the bright neighbor to the upper right subtracted. (B) Stack of the best-fit library PSF\null.  (C) The original minus the library PSF\null.  The subtraction is done in the {\tt \_flc} frame, then the pixels are stacked. (D) The original minus the optimized PSF\null.  (E) The previous panel with a neighbor added to the previous frame that has 10\% of the source brightness and offset by 42~mas.  (F)  The source+neighbor pixels fitted with a single-star PSF; the centroid moves up and to the right to account for the neighbor. 
\label{fig:faint_limit}
}
\end{figure*}

In order to place the most stringent limit possible on the lens brightness, we fit the above-referenced PSF to stars of similar brightness and color to that of the source, and adjusted the shape of PSF to optimize the fit (again, in the un-resampled frame). We saw a $\sim$1\% variation of the PSF, which is typical of minor, ``breathing''-related changes in telescope focus.  

Figure~\ref{fig:faint_limit} shows stacks of the source in our 2017 data at successive stages of analysis and illustrates the process of searching for photometric evidence of the lens.  In particular, panel (D) shows the subtraction with the improved PSF\null.  For illustration purposes, we show in panel (F) the residual pattern expected for a lens that is 10\% the brightness of the source.  There is no such signature visible in panel (D).  We quantify this as follows.

Using the improved PSF, we modeled the source image as a superposition of two stars separated by 42~mas (about 1 detector pixel) along the implied direction of proper motion\null.  We did this simultaneously in the $4\times 250$~s F814W images from 2017.  We found a best-fit flux for the lens of $-1$\% of the source flux.  We performed the same fits to similar-brightness neighboring stars, and found a distribution between $-2$\% and $+1$\%.  Finally, we added to each neighbor star a companion of 2.5\% at the presumed location, and were able to recover the added flux at a level of $2.5 \pm 1$\%.  We would clearly detect a lens if it were there.

We therefore conclude that the brightest the lens could be at the presumed location in 2017 is $\sim$1\% of the source flux. We also explored how bright a lens could be at an offset of 42 mas in {\it any\/} direction from the source.  We were able to rule out any lens brighter than 2.5\% of the source flux.

A luminosity of $\sim$1\% of that of the source corresponds to an apparent F814W or $I$ magnitude of $\sim$24.6. At the distance and extinction of the lens, the limit on its absolute magnitude is $M_I\gtrsim10.2$ (corresponding to a main-sequence star of about $0.2\,M_\odot$ or less, or a white dwarf of $\approx$$0.6\,M_\odot$ with a cooling age older than $\sim$$10^7$~years). 

These considerations leave a single BH---or a BH with a mass greater than $\sim$$3.55\,M_\odot$ in a binary paired with another BH, or with a NS or faint star---as the only viable possibilities for the lens.
\subsection{Constraints on Binarity}

Having established that the \OB\ lens is an extremely faint, or non-luminous, object with a mass of $7.1 \pm 1.3\, M_\odot$, we now consider in more detail whether the lens could be a binary system rather than a single, isolated BH\null. 

\OB\ was a high-magnification event, making its light curve especially sensitive
to lens binarity. As described above, the event was intensely monitored photometrically by several microlensing groups in search of
planetary-companion signatures, and no manifestation of a binary lens was seen. 

In fact, lens binarity will strongly affect both the photometric and the astrometric signature
if the angular separation between the components is around the angular Einstein radius, $\thetaE$, and it therefore has to be either much smaller or much larger. In either case, such a 
binary lens would produce a diamond-shaped caustic.

For a close binary, this caustic would be located around the center of mass of the lens,
and both the photometric and astrometric microlensing signatures would be strongly altered
if the source-lens trajectory passed over it or came close.
Specifically, for a mass ratio $q$ and a separation $d\,\thetaE$, the caustic extends to \citep{Schneider1986,Erdl1993,Dominik1999}
\begin{equation}
    s_\pm^\mathrm{c} = \pm \frac{2q}{(1+q)^2}\,d^2\,.
\end{equation}
Comparing this to $u_0 \simeq 0.00271$ yields the constraint $d < 0.07$ for the 
equal-mass case, with less than 11\,\% variation over the range $q \in [0.4,1]$.
With $D_\mathrm{L}\,\thetaE = 8.2~\mbox{AU}$, this implies a separation smaller
than $0.6~\mbox{AU}$\null. Considering a systematic noise floor of $\sim$5~mmag, deviations from a point-lens light curve exceed this for the given $u_0 \simeq 0.00271$ and blend ratio $g \simeq 20$ if $d > 0.022$, so that our data imply an upper limit to the separation of $\sim$0.18~AU.

In such a case, since one of the components has to have a mass of at least $3.55 \pm 0.55\,M_\odot$, the merging timescale for such a system is $\lesssim$$10^7$~years \citep{Carroll2006}.
Given that there is no evidence of active star formation
in this location, this seems unlikely.

In the case of widely separated binary components, the observed properties of the lens refer to one of them, leading to the same values as by assuming that the lens is a single isolated body, so that the mass determination of the putative BH remains valid. 

With $\thetaE$ referring to the mass of the object identified, in the wide-binary case
one finds the caustic extending to \citep[e.g.,][]{Dominik1999}
\begin{equation}
    s_\pm^\mathrm{w} = \pm \frac{q}{d^2} \, .
\end{equation}
Considering that any companion should be dark as well as the noise floor of our photometric data, we exclude any companion
more massive than 10\,\% of the identified lens  within $230~\mbox{AU}$.

To further quantify the limits on a binary companion to the lens, we performed a \citet{Rhie2000}-type analysis in which we simulate hypothetical binary-lens light curves with the same epochs and properties as the real data. We then fit those light curves with a point-lens model to determine the potential signal from a given lens companion. For this test, we use a subset of the data that provides good coverage of the overall light curve and that is sufficient to provide dense coverage of the peak, where a potential companion would have the greatest effect \citep{Griest1998}. Specifically, we use the OGLE and MOA survey data sets and the microFUN data from Auckland Observatory, Kumeu Observatory, Vintage Lane Observatory, the Wise Observatory 0.46m, and the Weizmann Institute. Setting the requirement that the $\chi^2$ due to the planet is $>$50, we find that the sensitivity limits can be approximated with a piecewise function
\begin{equation}
|\log(d_{\rm lim})|  = \left\{\begin{array}{ll}
  1.54 & \quad\mathrm{for}~\log q > -0.4 \\
   0.3413 \log q + 1.677 \\
   \multicolumn{2}{l}{\qquad\mathrm{for}~-4.8 < \log q \le -0.4} \\
    0.04  &\quad\mathrm{for}~\log q \le -4.8 \, .
\end{array}
\right.\,.
 \end{equation}
 Thus, for $q=0.1$, this analysis also excludes companions with separations $ 0.4 \, {\rm AU} \lesssim a_\perp \lesssim 180 \, {\rm AU}$.

\subsection{The Lens is an Isolated Black Hole}

To summarize this section, we have found that the lens that produced the \OB\ microlensing event has a mass of $7.1\pm 1.3\,M_\odot$. There are stringent limits on its optical luminosity, ruling out the possibility that the lens could be an ordinary star of that mass. Its mass is greater than possible for a white dwarf or NS, or for a binary pair of white dwarfs and\slash or NSs. Even if the lens is a binary, at least one of its components must still be a BH---but only under unlikely circumstances could it actually be a binary at all. In the rest of our discussion, we will assume that the lens is an isolated, single BH.

\section{Properties of the Black Hole}

Table~\ref{table:lensproperties} summarizes the inferred properties of the BH lens that produced the \OB\ photometric and astrometric microlensing event. In this section we discuss some properties of the lens and its Galactic environment.

\begin{deluxetable*}{lcc}
\tablecaption{Properties of the \OB\ Black Hole Lens \label{table:lensproperties} }
\tablehead{
\colhead{Property} &
\colhead{Value} &
\colhead{Sources \& Notes\tablenotemark{a}} 
}
\startdata
Mass, $M_{\rm lens}$                      & $7.1 \pm 1.3\, M_\odot$               & (1) \\ 
Distance, $D_L$                           & $1.58 \pm 0.18$ kpc                   & (2) \\
Einstein ring radius, $\theta_{\rm E}$    & $5.18\pm 0.51$ mas                    & (3) \\
Proper motion, $(\mu_\alpha, \mu_\delta)$ & $(-4.36\pm 0.22, +3.06\pm 0.66) \rm\,mas\,yr^{-1}$ & (4) \\
Galactic position, $(X,Y,Z)$              & $(-4, -1580, -45)$ pc        & (5) \\
Space velocities, $(V,W)$                    & $(+3, \, +40) \kms$                   & (6) \\
\enddata
\tablenotetext{a}{Sources \& notes: (1) This paper, \S\ref{subsec:mass}; (2)~This paper, \S\ref{subsec:lensdistance}; (3) This paper, Table~\ref{table:parameters_SC}; (4)~This paper, derived from model in Table~\ref{table:parameters_SC}; absolute proper-motion components in \Gaia\/ EDR3 J2000 frame; (5)~Galactic position relative to Sun: $X$ in direction of Galactic rotation, $Y$ in direction away from Galactic center, $Z$ perpendicular to Galactic plane toward north Galactic pole; (6)~Space-velocity components relative to Sun, assuming zero radial velocity: $V$ in direction of Galactic rotation, $W$ toward north Galactic pole; $U$ component toward Galactic center is undetermined due to unknown actual radial velocity.} 
\end{deluxetable*}

\subsection{Galactic Location} 

The BH lies in the direction of, but is considerably closer than, the Galactic bulge. The fifth row in Table~\ref{table:lensproperties} gives the rectangular components of its Galactic position relative to the Sun. The BH is located in the Galactic disk at a distance from us of $\sim$1580~pc toward the Galactic center, and about 45~pc below the Galactic plane. This places it approximately between the Scutum-Centaurus and Sagittarius-Carina spiral arms of the Milky Way, in the terminology of \citet{Reid2019}. There are no conspicuous star-forming regions, young stellar objects, or SN remnants at this location.

\subsection{The Surrounding Stellar Population} 

Most of the stars contained in the small FOV of our \HST\/ frames lie at considerably larger distances than the BH itself. In order to study a larger sample of the relatively nearby stellar population surrounding the spatial location of the BH, we selected stars from \Gaia\/ EDR3 that satisfy the following criteria: they lie within $20'$ of the BH on the sky, have trigonometric parallaxes with fractional errors of less than 10\% that fall in the range 0.55 to 0.70~mas, and are brighter than $G=17$. 

\begin{figure}
\begin{center}
\includegraphics[width=0.46\textwidth]{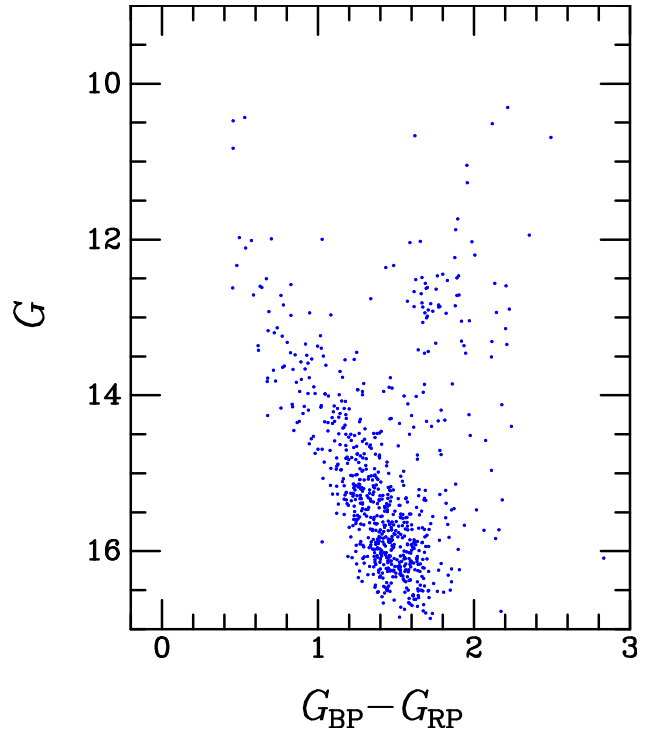}
\end{center}
\caption{
Color-magnitude diagram for a sample of stars in the Galactic disk at similar distances to that of the black hole, selected from \Gaia\/ EDR3 as described in the text. The population at this location contains main-sequence stars of a younger population, and older stars with redder colors. 
\label{fig:gaia_cmd}
}
\end{figure}

Figure~\ref{fig:gaia_cmd} shows the CMD of this population in the \Gaia\/ magnitude and color system. The CMD is afflicted by severe differential extinction, which blurs the stellar sequences. Nevertheless, it shows two prominent components: a moderately young and sparse population of main-sequence stars bluer than $G_{\rm BP}-G_{\rm RP}\simeq1.0$, and a denser older component of redder stars. There is little or no evidence for an extremely young population of massive stars, at the BH's location. 
However, with a transverse proper motion of $\sim 45 \kms$  (Table~\ref{table:lensproperties}), it would take only 0.2 Myr for the lens to traverse $20'$, so it's possible that the lens originated much farther away.

\subsection{Space Velocity, Population Constraints, Natal Kick}

The sixth row in  Table~\ref{table:lensproperties}  gives the components of the BH's space velocity relative to the Sun in the $V$ and $W$ directions (in the directions of Galactic rotation and toward the Galactic north pole, respectively). The third component of the space velocity, $U$, is undetermined since the radial velocity of the BH is unknown; changing the radial velocity to $-100$ or $+100\,\kms$ changes $V$ and $W$ only by about $\pm0.2$ and $\pm2.8\,\kms$, respectively. The relatively low value of $V$ indicates that the BH is not a member of an extreme Galactic-halo population, and may be moving in a roughly circular Galactic orbit (unless the radial velocity is large).  However, it does have a moderately high $W$ velocity perpendicular to the Galactic plane, so the orbit has a modest inclination to the Galactic plane. Apart from these considerations, there are few constraints on the age of the BH, except that it has likely had time to move far from its birthplace. 

\begin{figure}
\begin{center}
\includegraphics[width=0.46\textwidth]{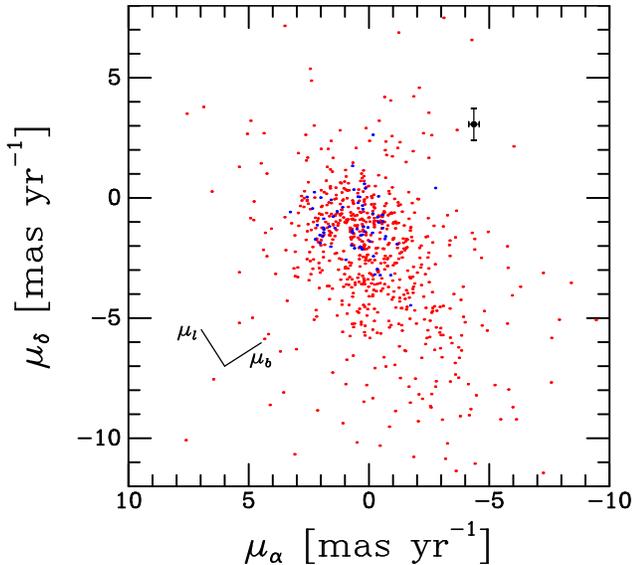}
\end{center}
\caption{
Proper-motion components from \Gaia\/ EDR3 for a sample of stars at similar distances to that of the black hole, selected as described in the text. Typical uncertainties are about $\pm$$0.05\rm\,mas\,yr^{-1}$ in both coordinates. Blue points represent a younger main-sequence population with \Gaia\/ colors of $G_{\rm BP}-G_{\rm RP}<1.0$, and red points an older and dynamically hotter population. The directions of Galactic longitude and latitude are indicated at the lower left. The black point with error bars shows the proper motion of the black hole, which is a significant outlier compared to the surrounding population. 
\label{fig:gaia_pm}
}
\end{figure}

Figure~\ref{fig:gaia_pm} plots the absolute proper motion of the BH (black filled circle with error bars, from the fourth row of Table~\ref{table:lensproperties}) and compares it with the proper motions of the neighboring \Gaia\/ stars from the sample described above. The directions of Galactic longitude and latitude are indicated at the lower left, showing that the dispersions are higher in the longitude direction than perpendicular to it. 
Members of the ``young'' population from Figure~\ref{fig:gaia_cmd} are plotted with blue points, and the older population with red points. The older population,
which has proper-motion dispersions of $\sigma_{\mu,l}=2.59\,\masyr$ and 
$\sigma_{\mu,b}=1.59\,\masyr$,
is dynamically hotter than the younger population, which has dispersions of $\sigma_{\mu,l}=1.08\,\masyr$ and 
$\sigma_{\mu,b}=1.05\,\masyr$.

The BH itself is a prominent outlier, predominantly in $b$, relative to the older and younger populations ($\sim$4 and $6\sigma$, respectively). This suggests it may have received a ``natal kick'' from the SN explosion associated with its birth---assuming that the BH arose from the surrounding population. Relative to the mean proper motion of the younger population, the BH's motion is offset by about $6.0\,\rm mas\,yr^{-1}$, corresponding to a tangential-velocity offset of $\sim$$45\,\kms$.  However, considering that its radial velocity is unknown, we cannot exclude the possibility that the BH might simply be passing through the surrounding population and could conceivably be significantly older. In a recent paper, \citet{Andrews2022} have given an extensive discussion of the possible natal kick of \OB.

\section{Discussion and Future Work}

\subsection{The Formation of Isolated Black Holes}

As reviewed in \S\ref{subsec:bhinbinaries}, it is generally believed that only
stars above $\sim$$20 \,M_\odot$ collapse to form BHs. To form an isolated
$\sim$$7\,M_\odot$ BH, such as the one discussed in this paper, either its
progenitor was a single star that experienced considerable mass loss through
stellar winds, it was produced through fallback in a weak SN explosion
\citep[e.g.,][]{Fryer1999}, or it was born in a close binary that lost mass
through a common-envelope mass ejection. In the binary scenario, the binary evidently
became unbound, presumably when its lower-mass companion itself became a
SN, and its mass loss and SN kick detached the binary.

The \OB\ BH's inferred mass of $\sim$$7.1\,M_\odot$ lies remarkably close to the
peak of the Galactic X-ray--binary BH mass distribution (see
\S\ref{subsec:bhinbinaries}). This may suggest a similar evolutionary path with
mass loss through close-binary  interactions, but a larger sample of mass measurements for
isolated BHs is needed to distinguish these different formation scenarios.

\subsection {Future Work}

There is no known X-ray or radio source at the position of \OB. 
Considering the high extinction toward the BH,
it is possible that it lies in a region of high density of interstellar matter.
If so, the accretion rate of material from the interstellar matter by the BH could conceivably be
large enough for it to be detected in deep X-ray and\slash or radio observations, as discussed in \S\ref{subsec:isolatedbhs}, and these would be worth attempting.
 
Finally, we note the potential of future facilities, such as the Very Large Telescope Interferometer (VLTI), Nancy Grace Roman Space Telescope, and the Rubin Observatory, in this area. The GRAVITY instrument with VLTI is currently
being upgraded to further improve its sensitivity \citep{Eisenhauer2019}, which will 
make it possible to measure $\thetaE$ from spatially resolved microlensed images 
\citep{Dong2019} of a large number of microlensed sources in the Milky Way.
 Roman will detect and characterize several thousand microlensing events \citep{Spergel2015, Penny2019}; with the same aperture size as \HST, its astrometric capabilities are expected to be similar \citep{JATIS2019}.  According to current plans, Roman will observe microlensing events over a $ \sim$2  square-degree region in the Bulge with a $\sim$15-minute cadence for several months each year; thus it will make thousands of photometric and astrometric observations of every long-duration event, leading to very high accuracy in both kinds of  measurements.  With such data, Roman should detect numerous BHs, yielding exquisite determinations of their masses, distances, and velocities, and providing important insights to our understanding of the formation and evolution of BHs. 
Rubin's wide-angle survey will provide deep, long baseline photometry and consistent data reduction, making it ideal for the discovery and photometric characterization of microlensing events \citep{Sajadian2019}, probing many different lines of sight (and hence stellar populations) across the Milky Way and Magellanic Clouds \citep{Street2018a}.  In the bulge, it will complement the Roman data by filling in gaps in the Roman survey cadence \citep{Street2018b, Strader2018}.  Rubin discoveries may need high-resolution timeseries imaging follow-up for full characterization of the events.

Note: After our paper was submitted, a study by \citet{Lam2022} that includes an independent investigation of \OB\ was posted on arXiv. We have not used any of the measurements or results from that paper in our analysis.

\acknowledgments

Based in part on observations made with the NASA\slash ESA {\it Hubble Space Telescope}, obtained at STScI, which is operated by the Association of Universities for Research in Astronomy, Inc., under NASA contract NAS~5-26555. Support for this research was provided by NASA through grants from STScI\null. \HST\/ data used in this paper are available from the Mikulski Archive for Space Telescopes at STScI,\footnote{\url{https://archive.stsci.edu/hst/search.php}} under proposal IDs 12322, 12670, 12986, 13458, 14783 and 15318.  

This work has made use of data from the European Space Agency (ESA) mission {\it Gaia\/} (\url{https://www.cosmos.esa.int/gaia}), processed by the {\it Gaia\/} Data Processing and Analysis Consortium (DPAC, \url{https://www.cosmos.esa.int/web/gaia/dpac/consortium}). Funding for the DPAC has been provided by national institutions, in particular the institutions participating in the {\it Gaia\/} Multilateral Agreement. 

The MOA project is supported by JSPS KAK-ENHI Grant Number JSPS24253004, JSPS26247023, JSPS23340064, JSPS15H00781, JP16H06287,17H02871 and 19KK0082.

We acknowledge
the help and dedication of the late Dr.\ John Greenhill who, as part
of the PLANET collaboration, played a key role in the efforts of collaboration in general, and in particular running
the Tasmania observatory operations.

K.C.S. spent some time at the European Southern Observatory, 
Institute for Advanced Study, and Harvard-Smithsonian Center for Astrophysics,
where some of this work was done, and thanks them for their hospitality.

{\L.W.} acknowledges support from the Polish NCN grant Daina No.\ 2017/27/L/ST9/03221.

J.-P.B. acknowledges support by the University of Tasmania
through the UTAS Foundation and the endowed
Warren Chair in Astronomy, and support
by ANR COLD-WORLDS (ANR-18-CE31-0002) at Le
Centre National de la Recherche Scientifique (CNRS)
in Paris and the Laboratoire d'astrophysique de Bordeaux. 

U.G.J. acknowledges funding from the European Union H2020-MSCA-ITN-2019 under Grant no.\ 860470 (CHAMELEON) and from the Novo Nordisk Foundation Interdisciplinary Synergy Programme grant no.\ NNF19OC0057374.

T.C.H. acknowledges financial support from the National Research Foundation (NRF; No.\ 2019R1I1A1A 01059609).

Y.T. acknowledges the support of DFG priority program SPP 1992 ``Exploring the Diversity of Extrasolar Planets'' (TS 356/3-1).




\facilities{HST (WFC3)}

\end{document}